\documentclass[submitting]{nst}
\usepackage{subfigure,dcolumn}
\usepackage[T2A,T1]{fontenc}
\usepackage[russian,english]{babel}
\usepackage{listings}
\usepackage{multirow}
\begin{document}
\title{Performance of plastic scintillator modules for top veto tracker at Taishan Antineutrino Observatory}
\author{Guang Luo}
\affiliation{School of Science, Sun Yat-sen University, Shenzhen 518107, China}
\author{Xiaohao Yin}
\affiliation{School of Physics, Sun Yat-sen University, Guangzhou 510275, China}
\author{Fengpeng An}
\email[Fengpeng An,]{anfp@mail.sysu.edu.cn}
\affiliation{School of Physics, Sun Yat-sen University, Guangzhou 510275, China}
\author{Zhimin Wang}
\email[Zhimin Wang,]{wangzhm@ihep.ac.cn}
\affiliation{Institute of High Energy Physics, Beijing 100049, China}
\affiliation{University of Chinese Academy of Sciences, Beijing 100049, China}
\author{Y.K.Hor}
\affiliation{School of Physics, Sun Yat-sen University, Guangzhou 510275, China}
\author{Peizhi Lu}
\affiliation{School of Physics, Sun Yat-sen University, Guangzhou 510275, China}
\author{Ruhui Li}
\affiliation{Institute of High Energy Physics, Beijing 100049, China}
\affiliation{University of Chinese Academy of Sciences, Beijing 100049, China}
\author{Yichen Li}
\email[Yichen Li,]{liyichen@ihep.ac.cn}
\affiliation{Institute of High Energy Physics, Beijing 100049, China}
\affiliation{University of Chinese Academy of Sciences, Beijing 100049, China}
\author{Wei He}
\affiliation{Institute of High Energy Physics, Beijing 100049, China}
\affiliation{University of Chinese Academy of Sciences, Beijing 100049, China}
\author{Wei Wang}
\email[Wei Wang,]{wangw223@sysu.edu.cn}
\affiliation{School of Physics, Sun Yat-sen University, Guangzhou 510275, China}
\affiliation{Sino-French Institute of Nuclear Engineering and Technology, Sun Yat-sen University, Zhuhai 519082, China}
\author{Xiang Xiao}
\affiliation{School of Physics, Sun Yat-sen University, Guangzhou 510275, China}

\begin{abstract}
The Taishan Antineutrino Observatory (TAO) experiment incorporates a top veto tracker (TVT) system comprising 160 modules, each composed of plastic scintillator (PS) strips, embedded wavelength shifting fibers (WLS-fibers), and silicon photomultipliers (SiPMs). This article highlights the performance of all produced modules following the production and readout/trigger design, providing insights for scintillation detectors with WLS-fibers. Three kinds of trigger modes and its efficiency have been defined to comprehensively evaluate the performance of this unique design, which has been verified for the batch production, along with comprehensive measurement strategies and quality inspection methods. In "module" mode, the detection(tagging) efficiency of the PS exceeds 99.67\% at a 30 photoelectron threshold, and even in "AND" mode, it surpasses 99.60\% at a 15 photoelectron threshold. The muon tagging efficiency meets TAO's requirements. The production and performance of the PS module set a benchmark for other experiments, with optimized optical fiber arrangements that enhance light yield and muon detection efficiency.
\end{abstract}



\maketitle

\section{Introduction}
\label{sec:intro}
For most neutrino or low background detectors, especially those near the ground\,\cite{Hakenmuller:2019ecb,TAO-CDR,SSAFSM,ZXJCYD,CFY,NEON:2022hbk,Singh:2017jow,Ang:2019ard}, a muon veto system is necessary to tag muon-induced particles going to the main detector\,\cite{NUCLEUS:2022vyj,JUNO:2023cbw,Li:2022wqc,Seo:2022vzr,W:Chen}. Cosmic ray (CR) muons can induce neutrons, producing gammas that mimic the coincidence signals. Therefore, efficient identification of muons to remove associated events is essential in the such experiments, which is the primary goal of the muon veto detector\,\cite{JUNO:2023cbw,Rico:2022aoc,Coveyou:2023fja,Adam:2007ex,Qain2021}. Plastic scintillation(PS) has been adopted as the basic unit of anti-coincidence detectors in many experiments\cite{WYP2021}, due to its advantages of easy machining, flexible structure design, stable performance, and good adaptability\,\cite{Moiseev:2007zz,Perotti:2006yc,Hao-Ran2021Discrimination,Zhuo2021Adaptability,ThaK:c,JND,YP:Chen,YT:Qu,WL2022,WuZ2018,HP2021,ps_ap}.

Taishan Antineutrino Observatory (TAO)\,\cite{TAO-CDR,T:CDR,TC:DR,TCD:R,Zhang:2024znq} scheduled to begin operation in spring 2025
, will independently measures the antineutrino energy spectrum of the reactor with unprecedented energy resolution for the
Jiangmen Underground Neutrino Observatory (JUNO)\,\cite{TAO-CDR,JUNO-CDR,JUNO:2015zny,JUNO:2021vlw,JUNOcpc}. TAO will provide a unique reference for other experiments and nuclear databases\,\cite{Capozzi:2020cxm}. A Top Veto Tracker (TVT) is designed serving as a part of the muon veto system, where the PS module is the key element of the TVT. A detailed description of the TVT is given in Ref\,\cite{Luo:2023inu}, where most aspects of the system design and optimizing process are covered. In short, it is a design with optimized WLS-fiber arrangement (uniformly bent in PS) that can improve light yield and muon detection efficiency.

This paper aims to describe the 
unique designs and performances.
We will introduce the PS module assembly and the production in sec.\,\ref{sec:Prod}. In sec.\,\ref{sec:System}, the testing system and the measurement results of one module will be presented from three main aspects: the light yield along its length of the PS module by the measurement of the CR muon, effective attenuation length, and coupling effect between wavelength shifting fiber (WLS-fiber) and silicon photomultipliers (SiPMs). In sec.\,\ref{sec:Perf}, details of the module performance will be discussed. In sec.\,\ref{sec:Dete}, the detection efficiency of the module is demonstrated under different threshold conditions. Finally, a summary is given in sec.\,\ref{sec:Sum}.

\section{Module production}
\label{sec:Prod}

\begin{figure*}[!htbp]
\begin{center}
\includegraphics[scale=0.85]{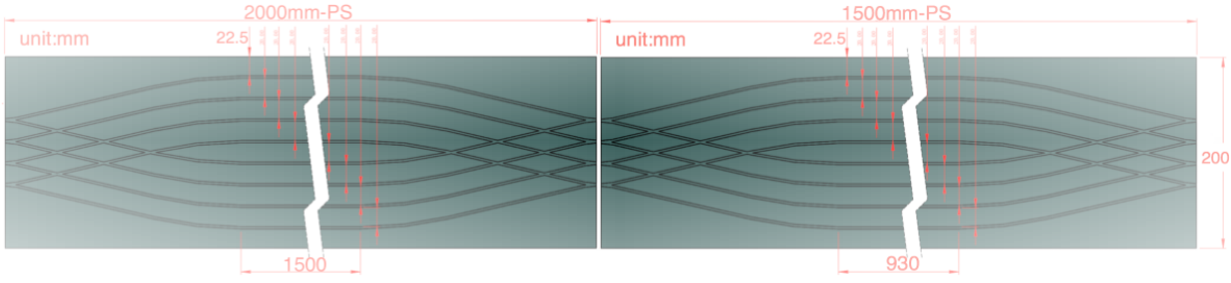}
\caption{The left panel is a design drawing of PS strips in a dimension of $2000\times 200\times20\, mm^3$ and The right panel is PS strips in a dimension of $1500\times 200\times20\, mm^3$.}
\label{fig:PSdesign}
\end{center}
\end{figure*}

The TVT is composed of 108 PS modules in a dimension of $2000(Length)\times 200(Width)\times20(Thickness)\, mm^3$ (2000mm-PS) and 52 PS modules in a dimension of $1500\times200\times20\, mm^3$ (1500mm-PS). Simulation optimization, prototype test and design of the PS strips for JUNO-TAO has been reported previously in Ref.\,\cite{Luo:2023inu,Lu:2023sbi,Min2023Performance}, and is demonstrated in Fig.\,\ref{fig:PSdesign}. Both 2000mm-PS and 1500mm-PS with eight 1.5\,mm WLS-fibers along its width direction (around 22.5\,mm spacing between neighboring fibers) which are laid and glued into the grooves on one surface of the PS. The eight fibers will be merged into four groups at each of the ends and coupled to the SiPMs.

Both 2000mm-PS and 1500mm-PS are fabricated by \textit{Beijing Hoton Nuclear Technology Co., Ltd.}. The PS modules are made with extruded plastic scintillator type of SP101 polymerized with liquid polystyrene added with P-triphenyl and POPOP. The WLS-fiber BCF92\,\cite{Tur:2009en,DietzLaursonn:2016tpy} with a diameter of 1.5\,mm is used and its end surface is flat. Two fibers are focused into a single group that can be coupled with optical sensors. To ensure the precise arrangement of WLS-fibers in the design, two methods were utilized: slotting and drilling. Firstly, a slot was made in the middle of the plastic scintillator (4\,mm deep), and holes were accurately drilled at both ends of the plastic scintillator. Subsequently, The groove was polished and cleaned. Finally, the WLS-fibers were placed at the bottom of the slot and secured, followed by filling with optical silicone grease (Type: SL600).

\begin{figure}
   \subfigure[]{
   \begin{minipage}[t]{0.45\linewidth}
   \centering
   \includegraphics[width=3.6cm]{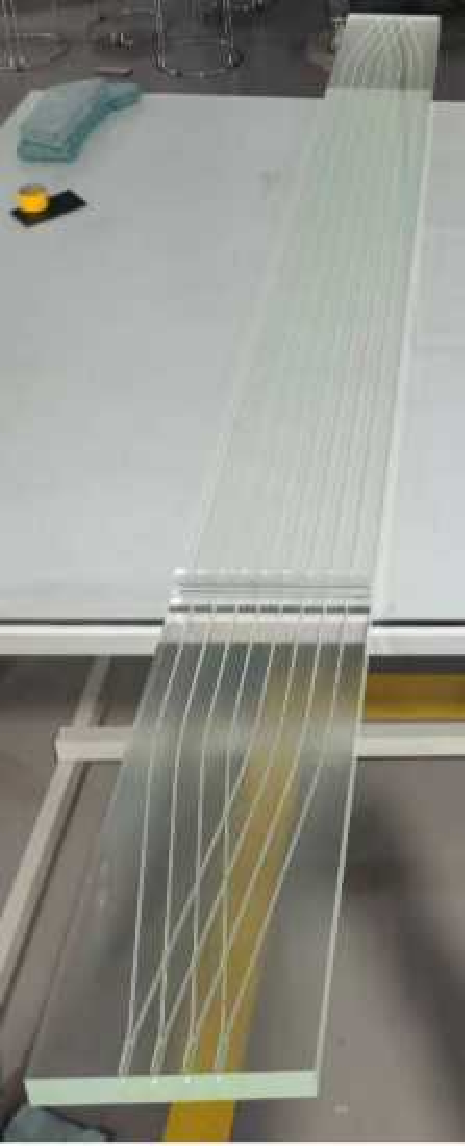}
    \label{fig:PS_physa}
    \end{minipage}
    }
    \subfigure[]{
    \begin{minipage}[t]{0.45\linewidth}
    \centering
    \includegraphics[width=3.7cm]{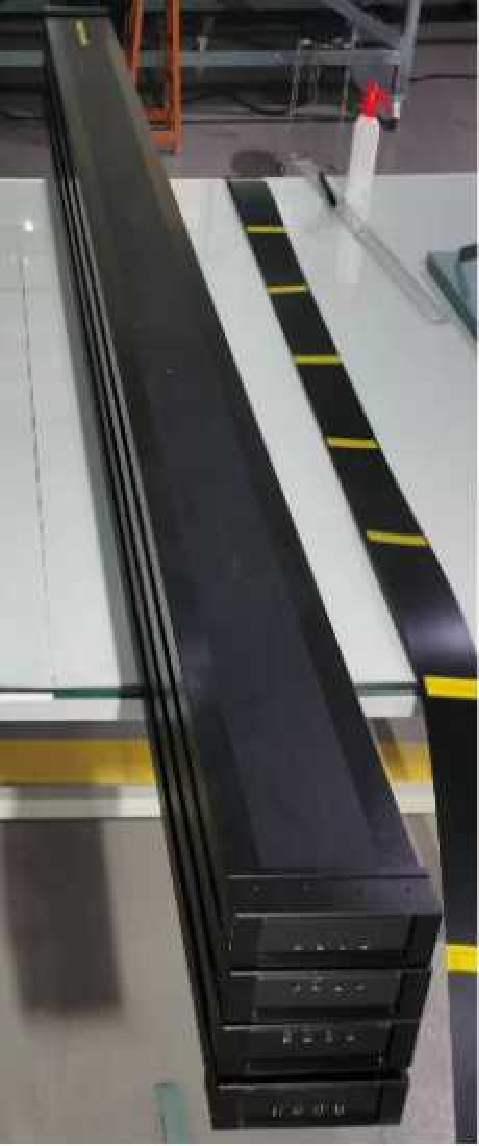}
    \label{fig:PS_physb}
    \end{minipage}
    }
    \caption{(a) The photo shows 2000mm-PS with inserted 1.5\,mm WLS-fiber. (b) The photo shows 2000mm-PS with aluminium film packaging and protective layer packaging on the periphery.}
    \label{12}
\end{figure}

Fig.\,\ref{fig:PS_physa} is the physical image of 2000mm-PS with inserted 1.5\,mm WLS-fibers, all surfaces are polished, and each fiber can be clearly seen in the arrangement of the PS.
The PS is first wrapped with 0.08\,mm aluminum foil, which serves as a reflective film, then with another 0.8\,mm PVC layer providing insulation and protection, and finally packaged with a black adhesive tape layer that blocks light.
Fig.\,\ref{fig:PS_physb} is the image of finished 2000mm-PS.

\begin{figure}
   \subfigure[]{
   \begin{minipage}[t]{0.95\linewidth}
   \centering
   \includegraphics[width=0.9\linewidth]{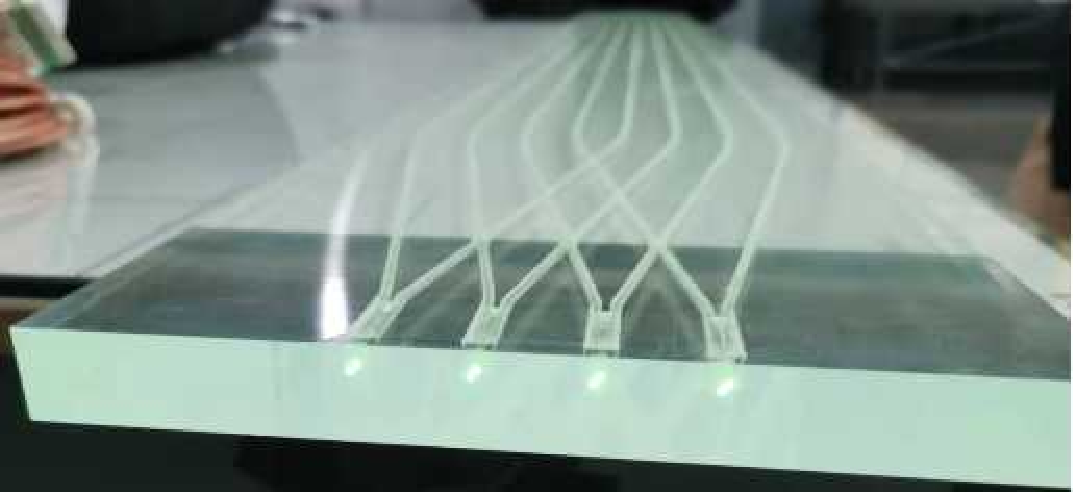}
    \label{fig:PS_physc}
    \end{minipage}
    }
    \subfigure[]{
    \begin{minipage}[t]{0.95\linewidth}
    \centering
    \includegraphics[width=0.9\linewidth]{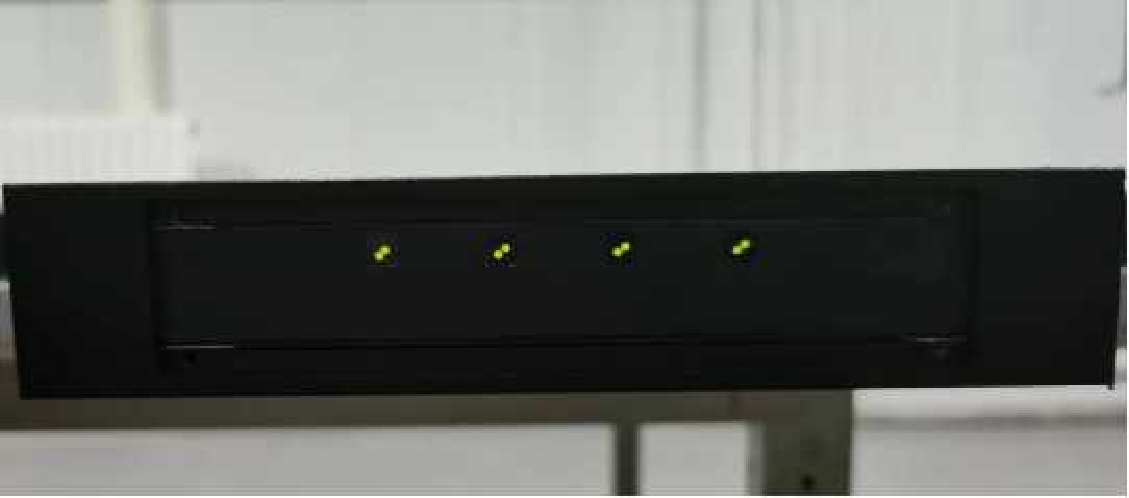}
    \label{fig:PS_physd}
    \end{minipage}
    }
    \caption{(a) The image shows 2000mm-PS backend with inserted 1.5\,mm WLS-fiber. (b) The image shows 2000mm-PS backend with Al film packaging and protective layer packaging on the periphery. }
    \label{34}
\end{figure}

As shown in Fig.\,\ref{fig:PS_physc}, the brightness of optical fiber and the PS can still be distinguished even without shading treatment. SiPM windows of dimension $4\times 4 mm^2$ have been cut on the backend of the covered PS. Fig.\,\ref{fig:PS_physd} shows the image of one end of the PS module, eight bright spots in four groups can be clearly seen from the $4\times 4 mm^2$ optical windows, which are the positions of the WLS-fibers.

\begin{figure}
   \subfigure[]{
   \begin{minipage}[t]{0.95\linewidth}
   \centering
   \includegraphics[width=0.9\linewidth]{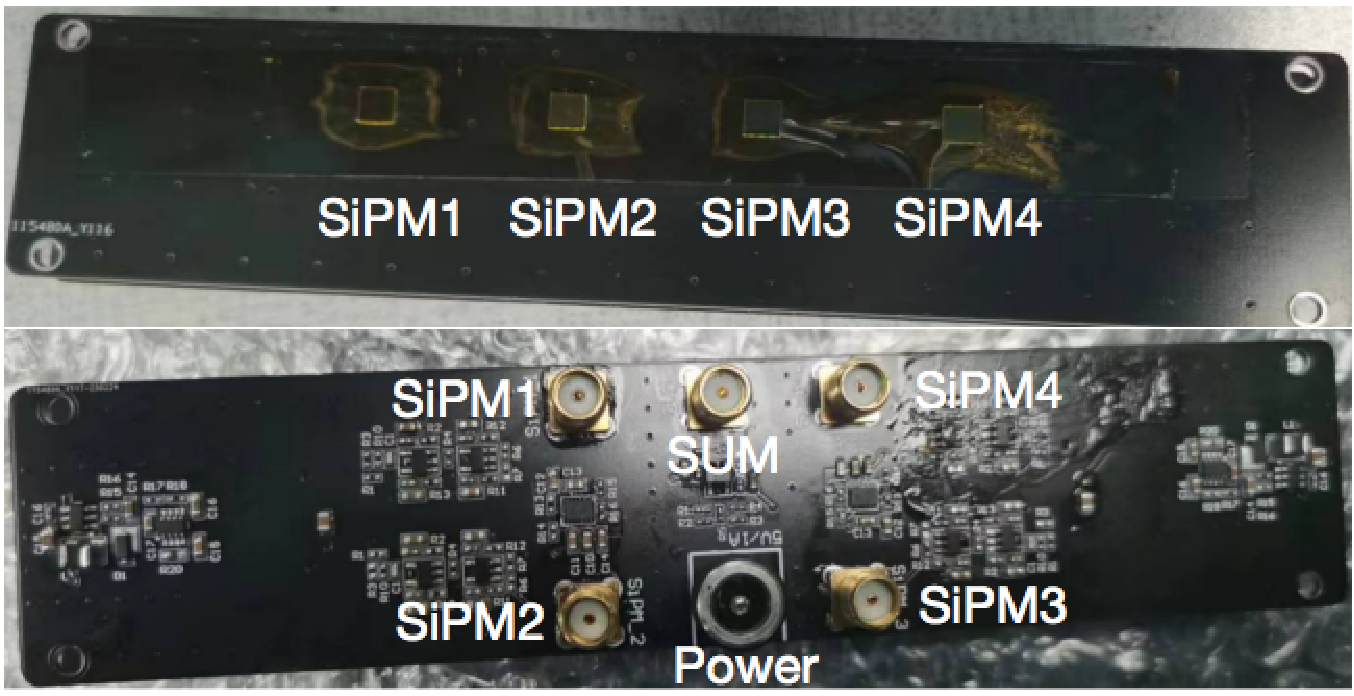}
    \label{fig:PS_physe}
    \end{minipage}
    }
    \subfigure[]{
    \begin{minipage}[t]{0.95\linewidth}
    \centering
    \includegraphics[width=0.9\linewidth]{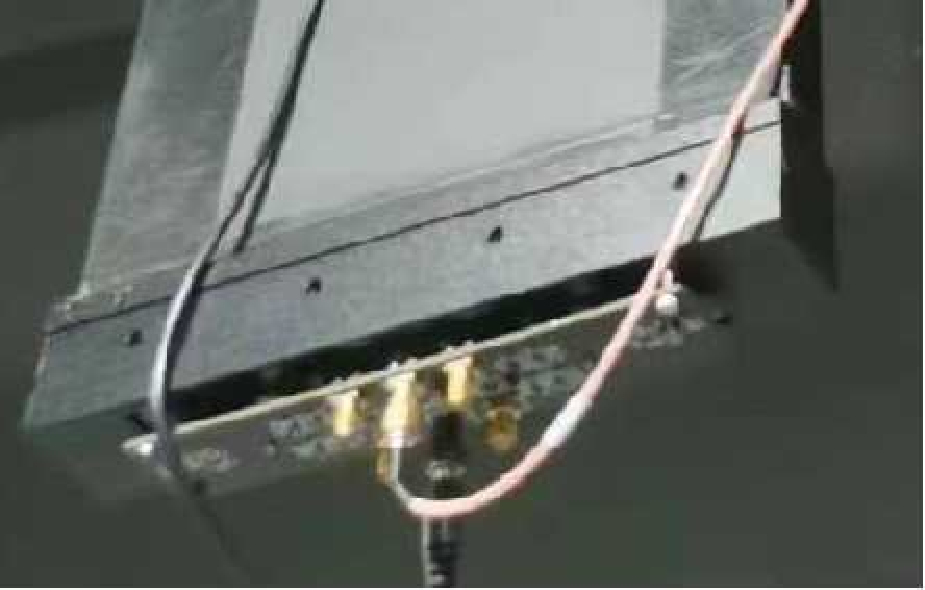}
    \label{fig:PS_physf}
    \end{minipage}
    }
    \caption{(a) The layout of 4 SiPMs on the PCB board and the structure diagram of the electronic board for signal readout. (b) The physical image after the backend has been connected to electronics.}
    \label{fig:ps}
\end{figure}

Four SiPMs of J-series MicroJ-40035-TSV with integrated readout are used as the photon sensor at each end of the PS module\cite{LW2021,TA2023,Sipm_ps,Si:SiPM}. Fig.\,\ref{fig:PS_physe} shows the layout of SiPMs on a PCB and the circuit of the output signals channel. Four SiPMs (SiPM1, SiPM2, SiPM3, and SiPM4) are coupled to the optical windows, the designed PCB can read-out each SiPM signal and also sum up the four SiPM signals. According to previous simulation results from Ref\,\cite{Luo:2023inu}, the difference among the four SiPMs on one end are very small. The design of reading out each SiPM signal is to measure and verify precisely the difference among the four SiPM signals. The design of four SiPM summing channels is to reduce the number of readout channels, and the final design of the PCB will only have one summing channel output. At present, the design of this type of PCB is mainly for module quality inspection and performance testing. The current version of PCB are fabricated by \textit{Dual Rainbow Technology Co., Ltd.}\,\cite{dualrainbow}. Fig.\,\ref{fig:PS_physf} is the photo of the backend of the PS module during mass testing after the PCB is mounted. The aluminum fixtures are used to mount the PCB onto the plastic scintillator. After the installation of the fixture, they are packaged as a whole and finally coated with black epoxy for light-tight treatment. The output signal line uses coaxial cables with SMA connectors.

\section{Testing system and measurements}
\label{sec:System}

\begin{figure*}[!htbp]
\begin{center}
\includegraphics[scale=1]{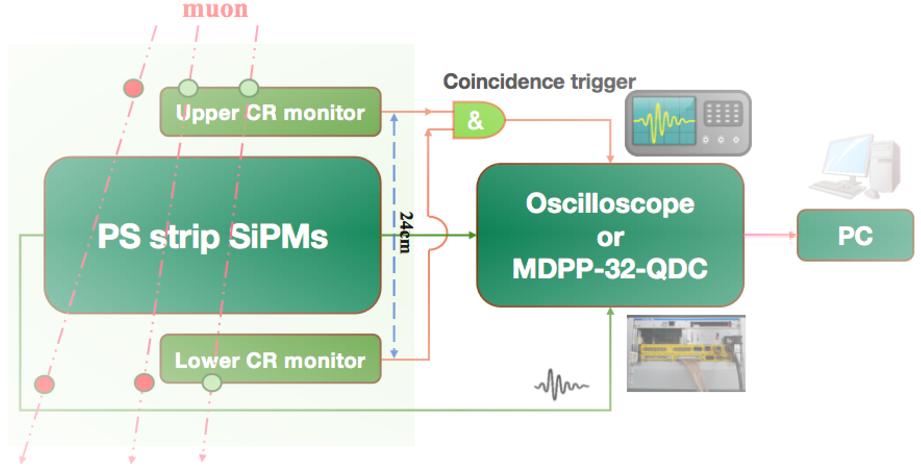}
\caption{
A flowchart of the data flow are used to measure the performance of each PS strip by muon at 9 locations.
}
\label{fig:testsys}
\end{center}
\end{figure*}

Fig.\,\ref{fig:testsys} is a flowchart of the data flow used to measure the performance of each PS strip by muon at 9 locations. A CR muon monitor system consists of two plastic scintillators with the type of EJ200\,\cite{cite:Ej200} with dimensions of $6\times 6\times 18\, cm^3$ and a 2-inch Photo Multiplier Tube (PMTs)\,\cite{Chen2023,YF2011} of type XP3232. The coincidence of the two monitors is used as a trigger to identify a muon passing through the 2000mm-PS, and to characterise the performance of the strip. The vertical distance between the two monitors is 24\,cm to tag muons going through the 2000mm-PS strip. In Fig.\,\ref{fig:testsys}, the two CR monitors are placed in the center of the PS module. A high-resolution oscilloscope with type lecroy-HDO4104A is used to obtain a total of four channels of signals, including the two of the CR monitors, and another two of the summation or single channels of the SiPMs attached to the two ends of the PS strip. In addition, another fast high-resolution time and amplitude digitizer with type MDPP-32-QDC with less dead time is also used in parallel for comparison, which is internally realized as a 32-channel adjustable low noise amplifier and a variable differentiation stage, followed by band pass filters and 80\,MHz sampling ADCs. MDPP-32-QDC can realize about eight times more channels than the used oscilloscope. For subsequent measurements, we can perform 8 PS strips measurements in parallel, and simultaneously measure the single and summed channels at the same time. 

\begin{figure*}
   \subfigure[]{
   \begin{minipage}[t]{0.45\linewidth}
   \centering
   \includegraphics[width=7.0cm]{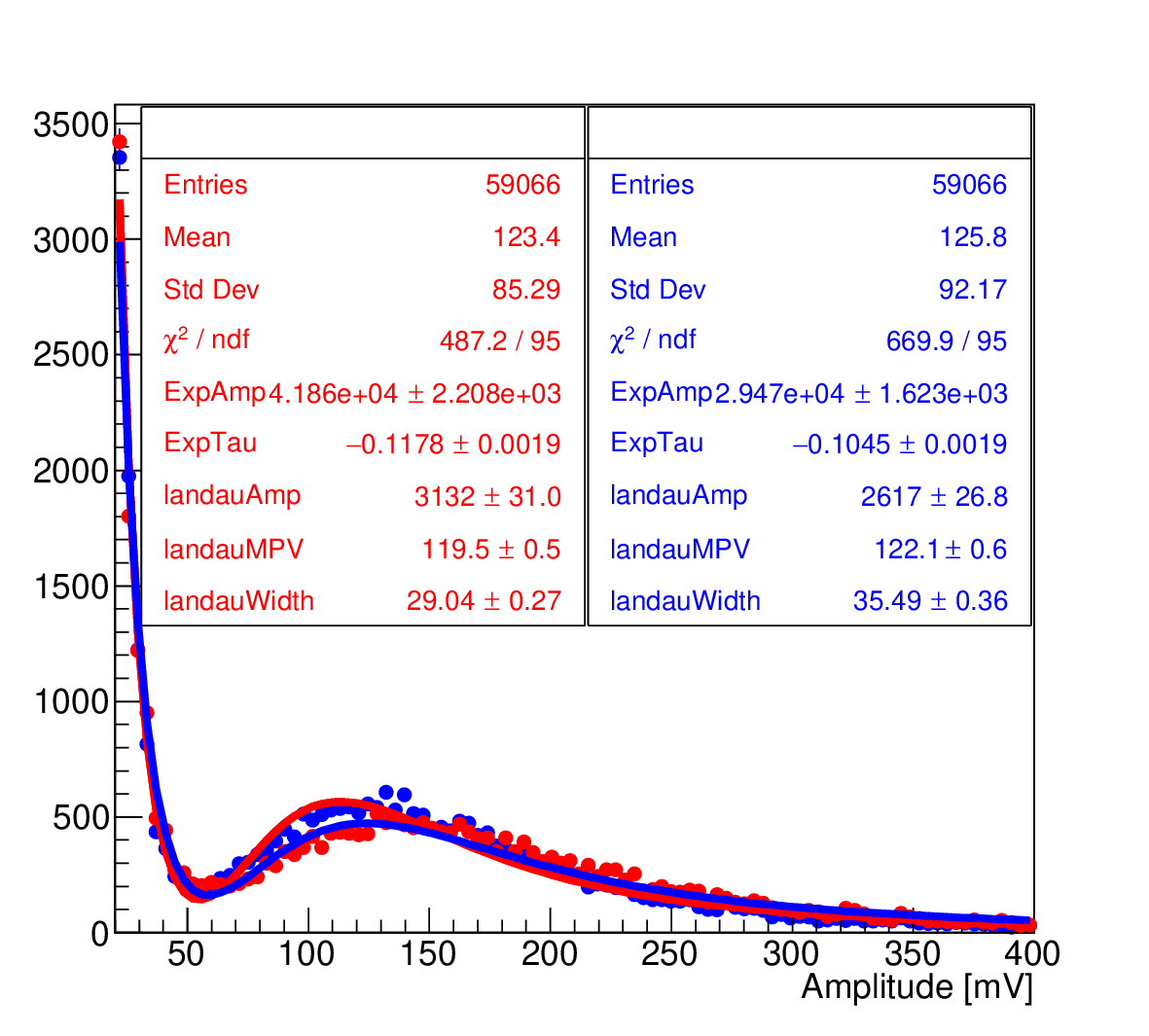}
    \label{fig:RestS1}
    \end{minipage}
    }
    \subfigure[]{
    \begin{minipage}[t]{0.45\linewidth}
    \centering
    \includegraphics[width=7.0cm]{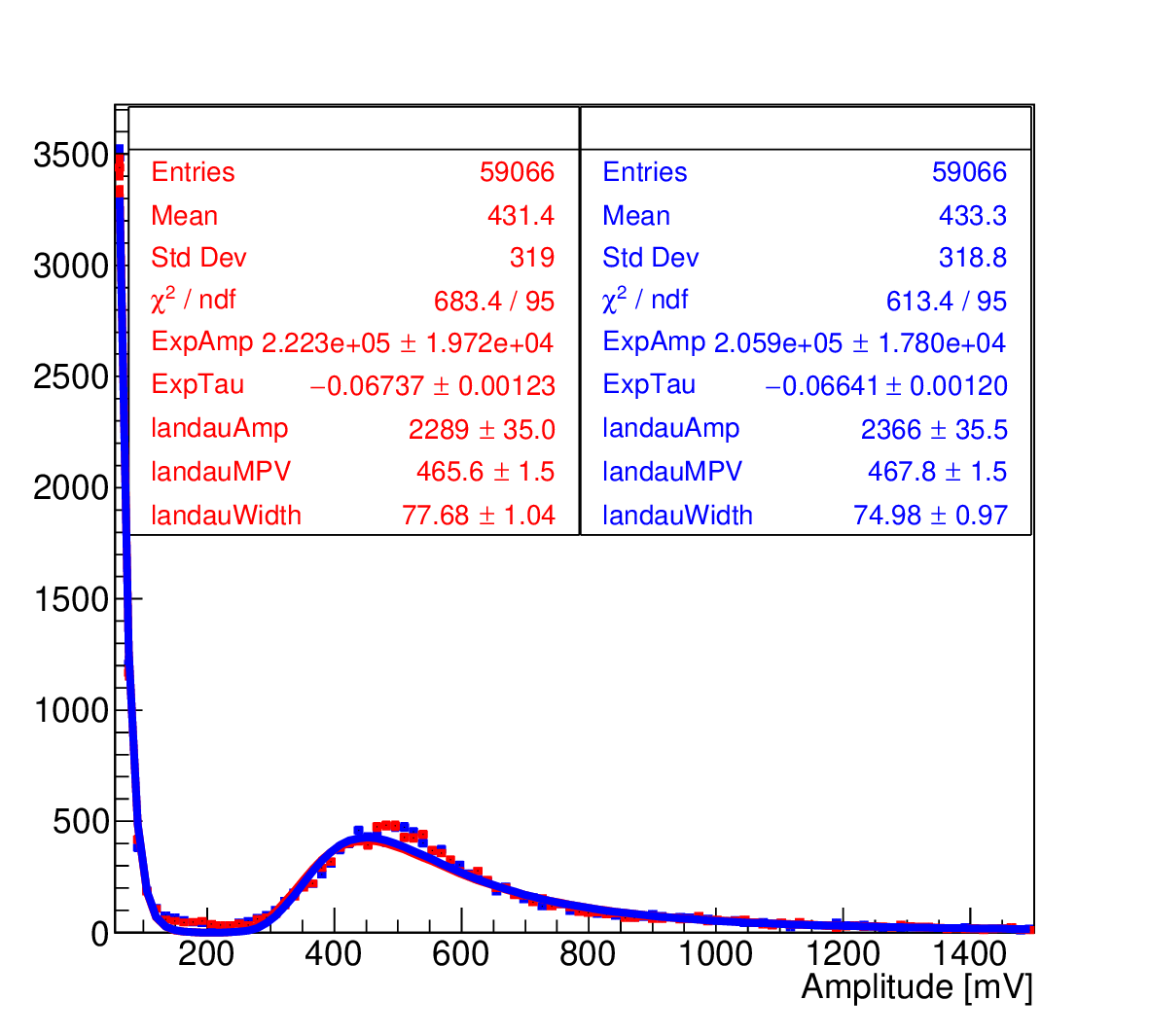}
    \label{fig:RestS2}
    \end{minipage}
    }
    \subfigure[]{
    \begin{minipage}[t]{0.45\linewidth}
    \centering
    \includegraphics[width=7.0cm]{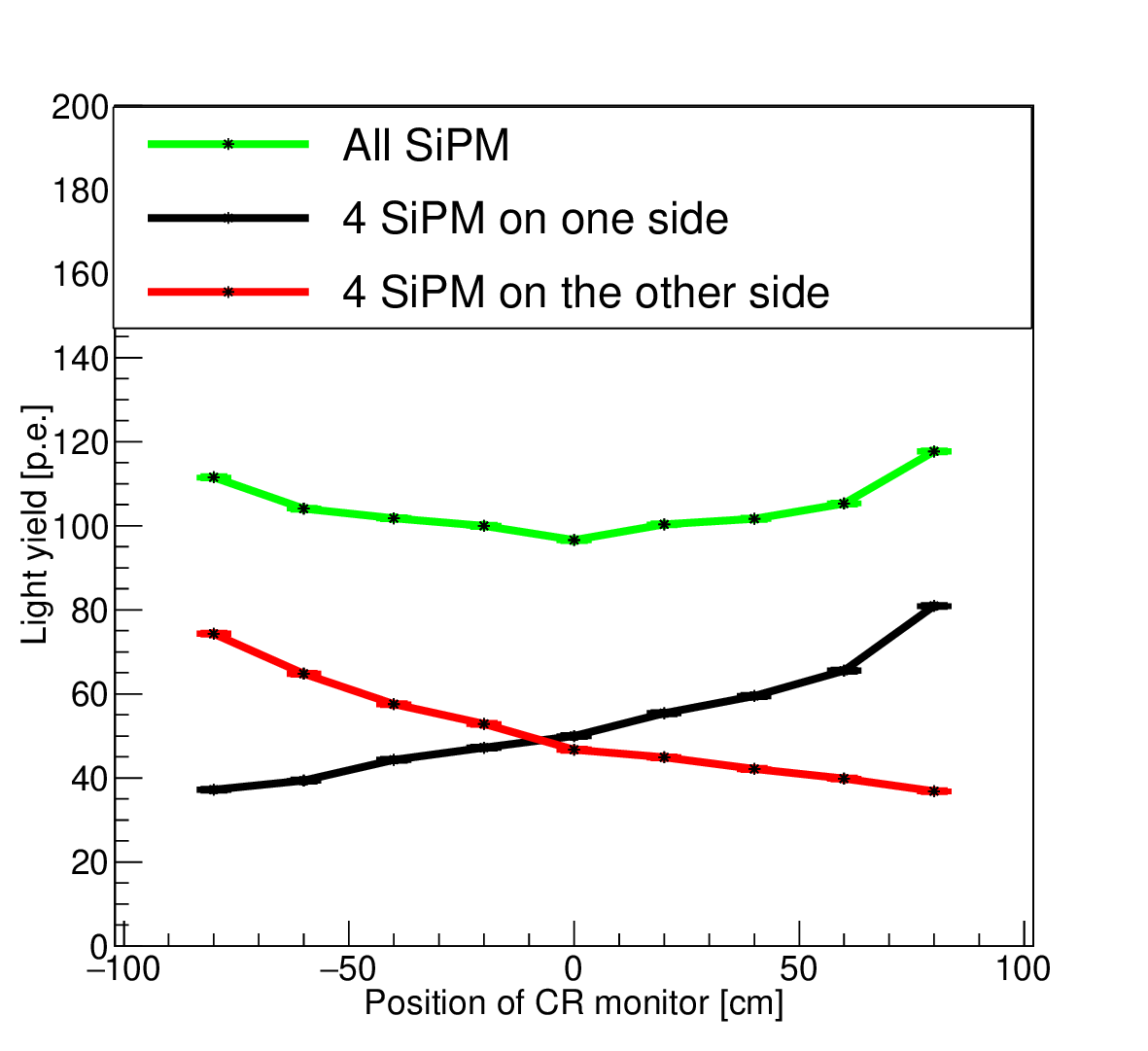}
    \label{fig:RestS3}
    \end{minipage}
    }
    \subfigure[]{
    \begin{minipage}[t]{0.45\linewidth}
    \centering
    \includegraphics[width=7.0cm]{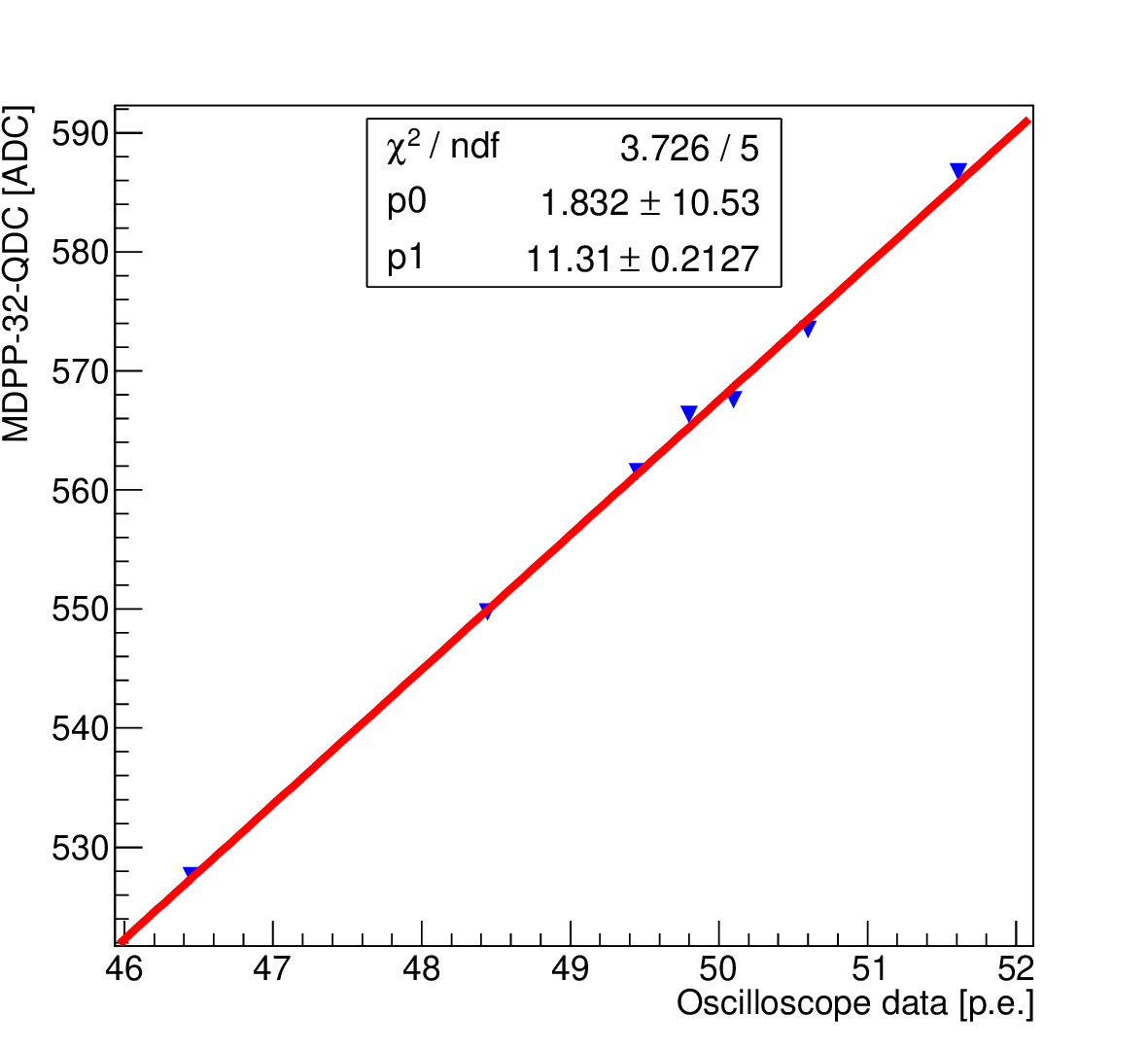}
    \label{fig:RestS4}
    \end{minipage}
    }
    \subfigure[]{
    \begin{minipage}[t]{0.45\linewidth}
    \centering
    \includegraphics[width=7.0cm]{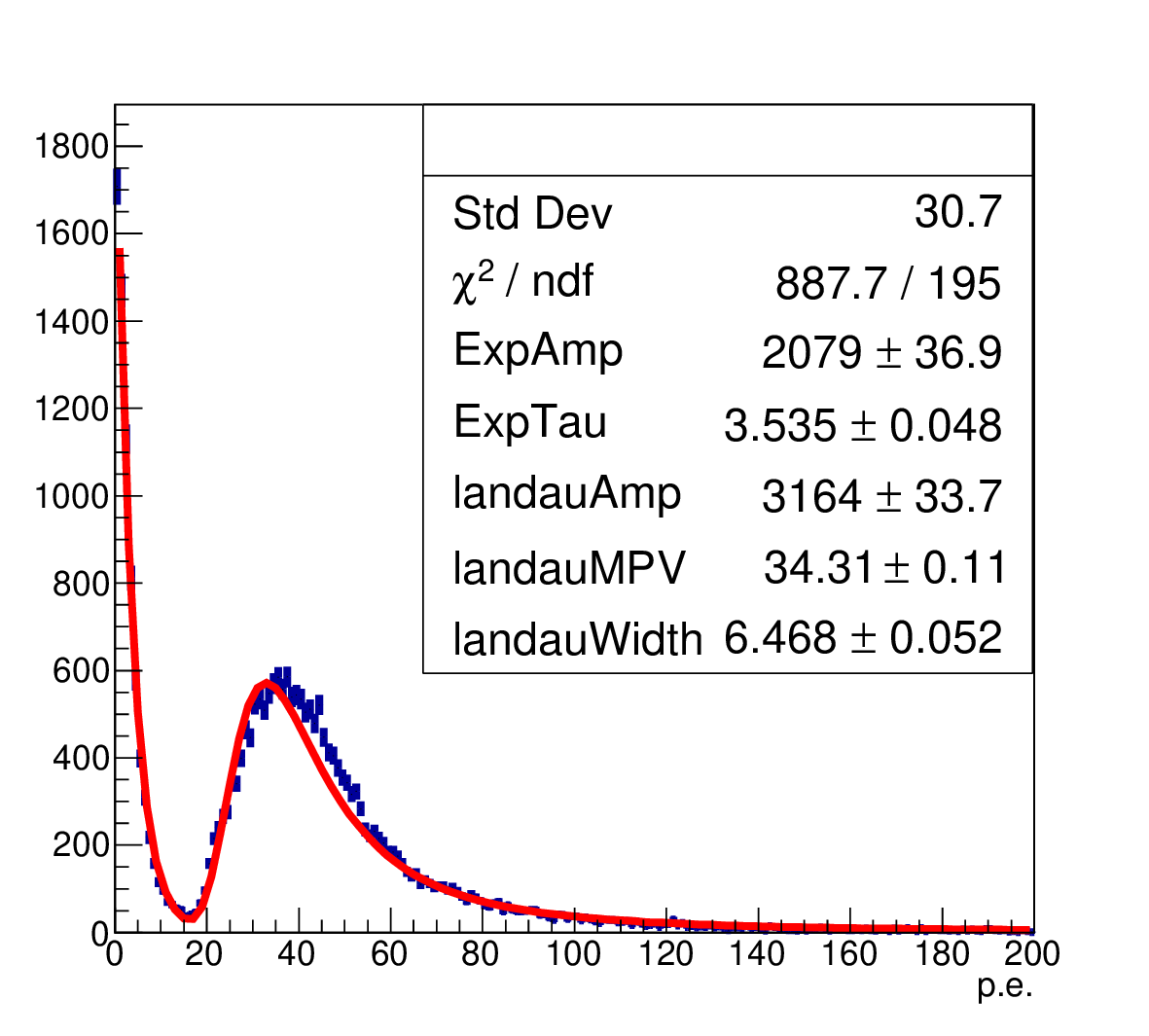}
    \label{fig:RestS5}
    \end{minipage}
    }
    \subfigure[]{
    \begin{minipage}[t]{0.45\linewidth}
    \centering
    \includegraphics[width=7.0cm]{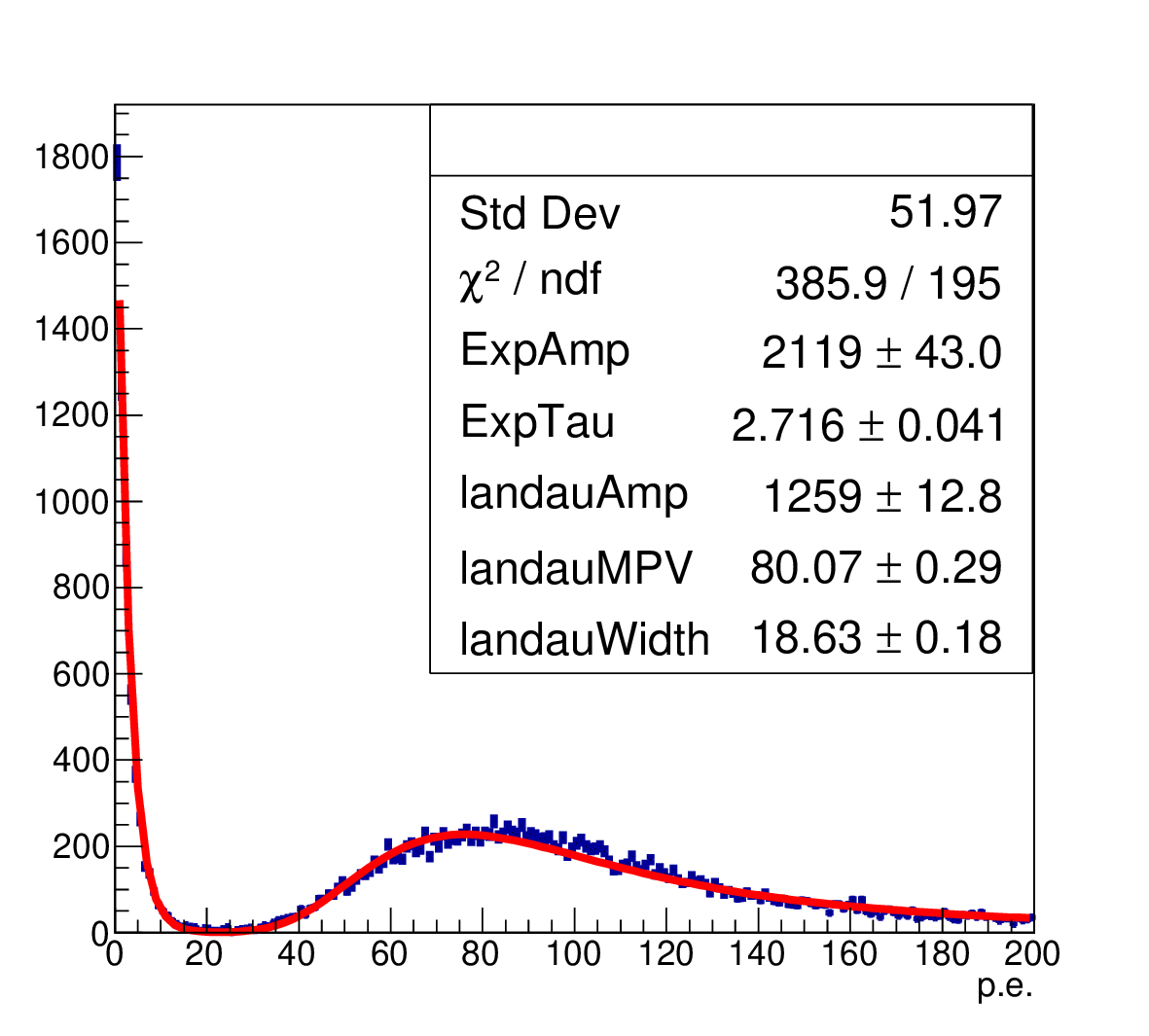}
    \label{fig:RestS6}
    \end{minipage}
    }
    \caption{(a) The amplitude spectrum of two adjacent SiPMs at one end. (b) The amplitude spectrum of the summation channels at both ends. (c) The distribution map of the light yield of a 2000mm-PS strip with the CR monitors at different locations. (d) MDPP-32-QDC data vs.\,Oscilloscope data. (e) The signal spectrum of the PS on one side when the CR monitor is located at -97\, cm(the outermost edge).  (f) The signal spectrum of the PS on the other side when the CR monitor is located at -97\, cm.}
    \label{35}
\end{figure*}

In the test, we warmed up the PMTs of the CR monitors and set their operating voltages both to -1020\,$\text{V}$. With a pre-test and calibration, the threshold for both PMTs of the CR monitors was set to -14\,$\text{mV}$ for muon identification. 

Fig.\,\ref{fig:RestS1} shows the distribution of the amplitude of the signal from two adjacent SiPMs on the same PCB at one end of the PS. The blue and red dots represent experimental data from SiPM1 and SiPM2, respectively. Except for a few datas that do not overlap, the two spectrum almost overlap in other areas, The spectrum is fitted by a joint function of exponential and Landau distribution for gamma and muons. For SiPM1, the most probable value (MPV) of signal amplitude for muon is 119.5\,$\text{mV}$. For SiPM2, the MPV of signal amplitude for muon is 122.1\,$\text{mV}$. The difference between the SiPM1 and SiPM2 is < 3\,$\text{mV}$, which is consistent with the previous simulation\,\cite{Luo:2023inu}, where
the signals from different WLS-fibers are relatively equal at the 5\% level. This also indicates that our WLS-fibers layout is effective and fairly uniform.
Fig.\,\ref{fig:RestS2} shows the distribution of the amplitude of the signal from the summation channel at both ends of the same PS when the CR monitors located at its center, the blue and red dots represent experimental data from one end and another end, respectively. It is obvious that these two spectra almost overlap. At one end, the MPV of the fitted Landau distribution is 465.6\,$\text{mV}$, at another end, the MPV of the fitted Landau distribution is 467.8\,$\text{mV}$, which are same and is almost four times of an individual SiPM channel as in Fig.\,\ref{fig:RestS1}.

To characterize the signal strength when the muon going through the plastic scintillator strips, the MPV fitted by Landau distribution can be used as an estimation of the light yield of the strip after converted into p.e., which will be measured and calculated at nine different locations along the longitudinal direction of the PS module. Fig.\,\ref{fig:RestS3} is measured results, where the x-axis is the position of the CR monitor in the longitudinal direction of the PS strip, and the error bar in x-axis represents the 6\,cm size width of the CR monitor. The y-axis is the measured light yield in p.e.\,units, which is calculated by converting the relationship between MPV and the signal amplitude of a single photoelectron with a pre-calibrated factor (10\,mV per p.e., claimed efficiency 38\% under the bias voltage 28\,V(the overvoltage is 3.5 V), cross talk ration 11.6\%). The black, red, and green lines represent the light yield of the 4 SiPM summation channel of one end, the other end and the sum of both ends, respectively. When the CR monitor is located in the range of -80 or 80\,cm of the PS strip, the minimum light yield can still be higher than 35 p.e.\,for single end, and the maximum light yield can almost reach 80 p.e.\,at single end. 

For a cross-check, we also compared the results from the two sets of data acquisition systems (oscilloscope or MDPP-32-QDC). Placing the CR monitors at the center of the PS strip, we measured seven PS strips using the two systems, and the MPV of each PS strip was fitted and obtained. In Fig.\,\ref{fig:RestS4}, the x-axis is the MPV from oscilloscope (in p.e.), and the y-axis is the MPV from MDPP-32-QDC (in ADC). The seven blue inverted triangle points are the MPVs of the seven PS strips, and fitted by a linear function. The red line is the fitted results. It is around 11.3\,ADC/p.e., represents the average ADC count value corresponding to each photoelectron. In addition, we sampled and measured the outermost position, placing the CR monitor near -97\,cm from the center of the PS strip. Fig.\,\ref{fig:RestS5} shows the signal spectrum of one end of the PS that is far away from the CR monitor, the MPV of the fitted Landau distribution was 34.31\, p.e., Fig.\,\ref{fig:RestS6} shows the signal spectrum of the other end of the PS, which is near the CR monitor, the MPV of the fitted Landau distribution was 80.07\, p.e., similarly, from the comparison between Fig.\,\ref{fig:RestS5} and Fig.\,\ref{fig:RestS6}, it can be observed that the larger the MPV, the wider the Landau width. The width of the far end fitting is reduced by almost three times compared to the near end fitting width.

From the green line in Fig.\,\ref{fig:RestS3}, it can be seen that when the CR monitors are at the centre of the PS strip, the summed signal strength is the lowest and about 95\,p.e. The more you move the CR monitors towards the end, the higher the total light yield. When the CR monitors are out of the range of -80 or 80\,cm of the PS strip, the summed signal intensity is about 115 p.e. In order to characterize the attenuation of light generated by muons in the PS module, 
\begin{equation}
Y = Y_0 e^{-\frac{L}{L_0}};
\label{equ:0}
\end{equation}
We can use Formula.\,\ref{equ:0} to fit the points in Fig.\,\ref{fig:RestS3}. where $L_0$ is the effective attenuation length of PS module, $Y_0$ is the initial light yield. $Y$ and $L$ are the light yield and the position of the corresponding CR monitor in Fig.\,\ref{fig:RestS3}, respectively.
\begin{figure}[!htbp]
\begin{center}
\includegraphics[scale=0.4]{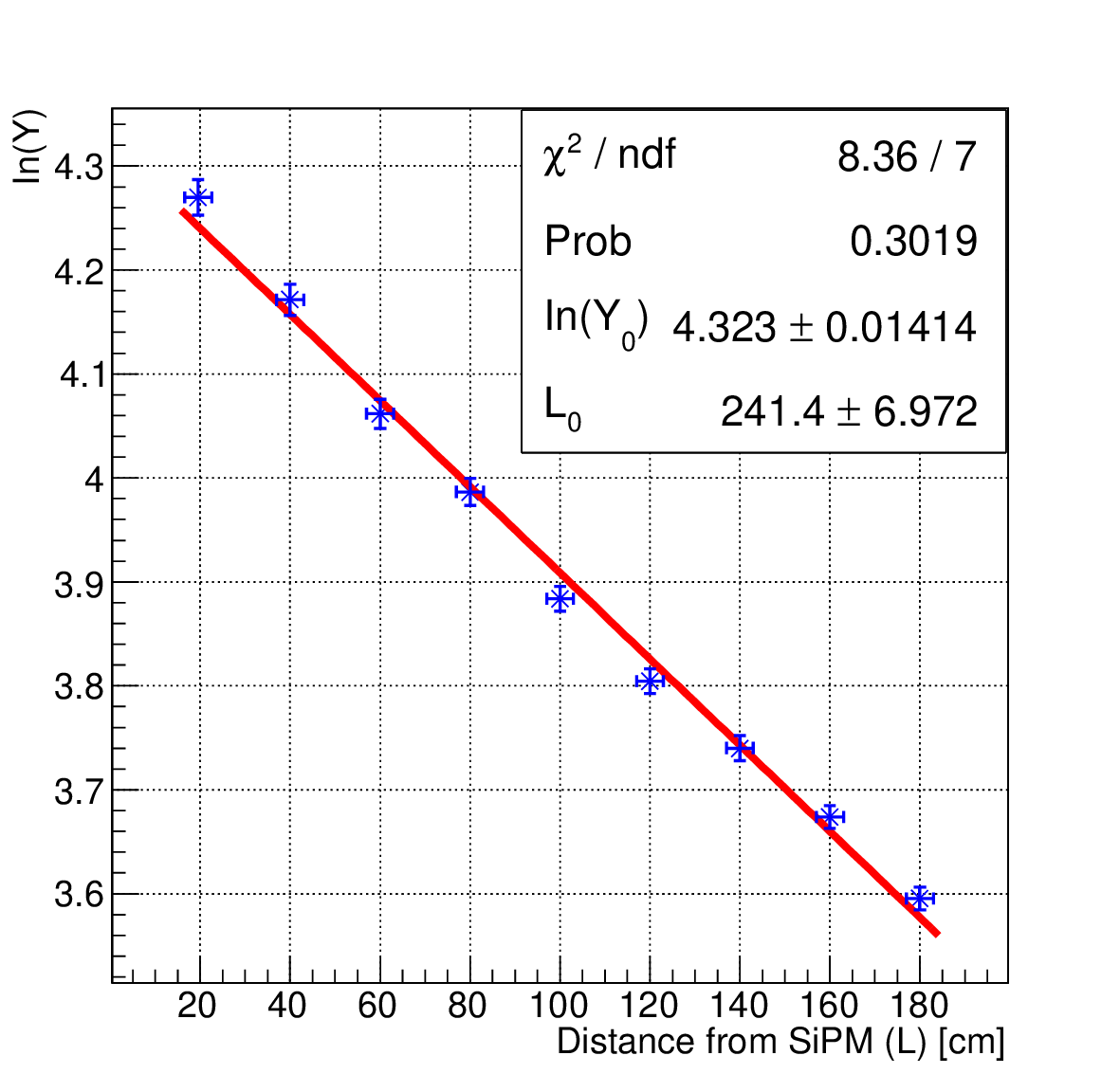}
\caption{The fitting result plot}
\label{fig:fitresult}
\end{center}
\end{figure}
In Fig.\,\ref{fig:fitresult}, by fitting the logarithm of light yield with distance from SiPM\cite{Yang:2022awv,YXW2020,GL2013}.
we obtained an effective attenuation length of 241 ± 6.97 cm and it met our design requirement of > 200 cm. According to simulations and some experiments, the intrinsic attenuation length of PSs is approximately 2 meters at a photon wavelength of 410 nm, while the intrinsic attenuation length of WLS-fibers is approximately 3.8 meters at a photon wavelength of 410 nm. The effective attenuation length is greater than the intrinsic attenuation length of the PS, indicating that the optimized arrangement of fiber insertion is effective.
It is known that the diameter of WLS-fibers is very small (1.5mm) compared to PSs, with a total volume of less than 5\%, and most photons are still attenuated in PSs. Therefore, the effective attenuation length is still mainly caused by the intrinsic attenuation length of PS.

\section{Module performance}
\label{sec:Perf}

\begin{figure}[!htbp]
\begin{center}
\includegraphics[scale=0.40]{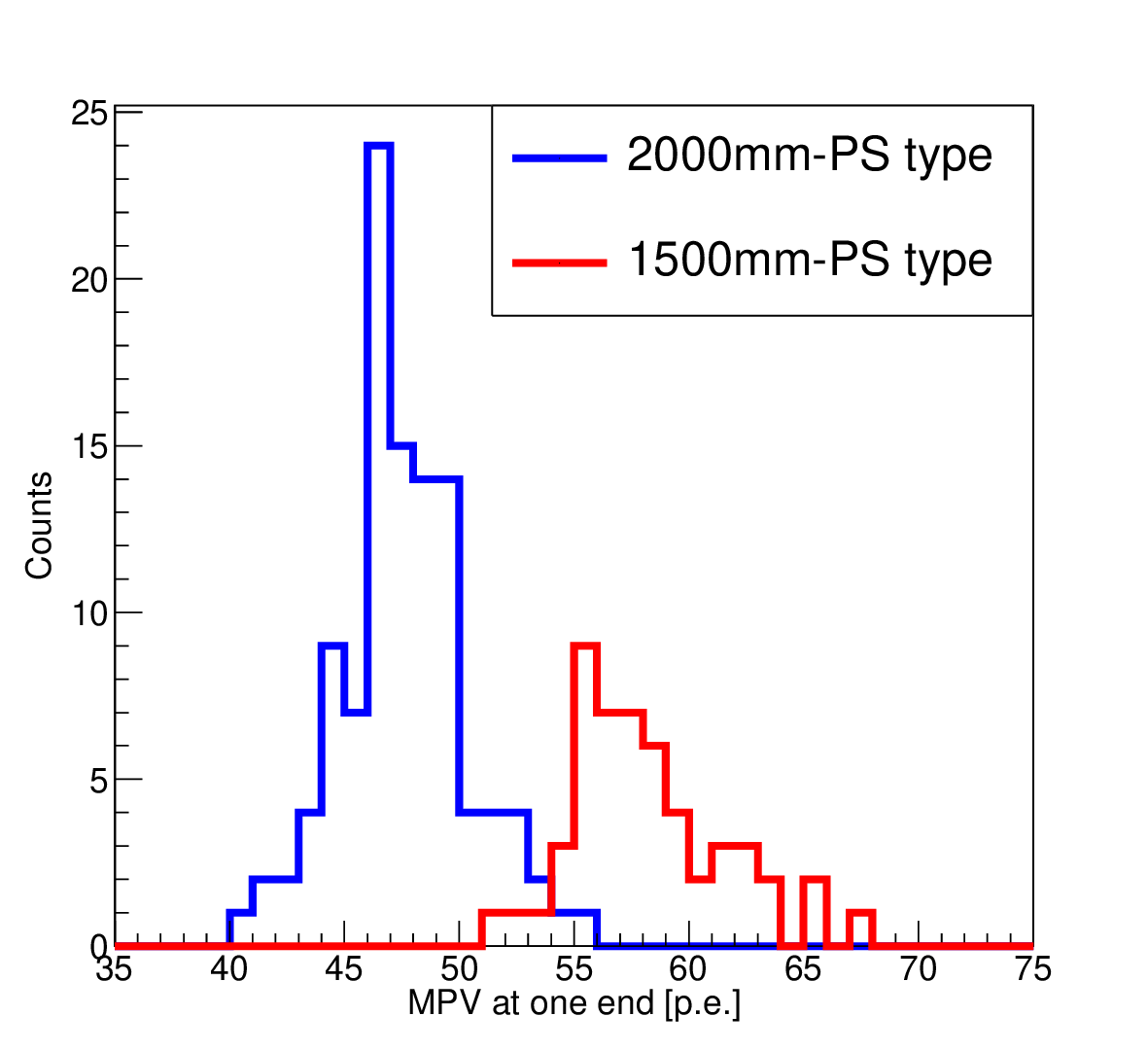}
\caption{MPVs of 108 2000mm-PS modules and 52 1500mm-PS modules.}
\label{fig:AllMPVs}
\end{center}
\end{figure}

We conducted four batches of acceptance and performance tests following the production in total. Fig.\,\ref{fig:AllMPVs} shows the MPV of single end of all the plastic scintillator strips when the CR monitors are located at the centre of the PS module. It can be seen in Fig.\,\ref{fig:AllMPVs} that the lowest light yield from single end of the 108 strips of 2000mm-PS modules (blue line) is about 40.8\,p.e., and the mode is measured to be 46\,p.e. It can be obtained from the red line for the 52 strips of 1500mm-PS strips, the lowest light yield of single end is about 51.5\,p.e., and the mode concentrates at 55\,p.e..

In addition, we tried EJ-550~\cite{YY550} silicone optical grease for the coupling between SiPM and WLS-fiber, rather than the coupling by only air during the test. Subsequent measurements were conducted to demonstrate that the coupling between SiPM and WLS-fiber by optical grease can increase the measured light yield by at least 5\,p.e.\,at one end (around 12.5\% in light yield for 2000mm-PS modules). This indicates that the light yield of our officially operated PS module will be even higher.

\section{Detection efficiency}
\label{sec:Dete}

\begin{figure}[!htbp]
\begin{center}
\includegraphics[scale=0.42]{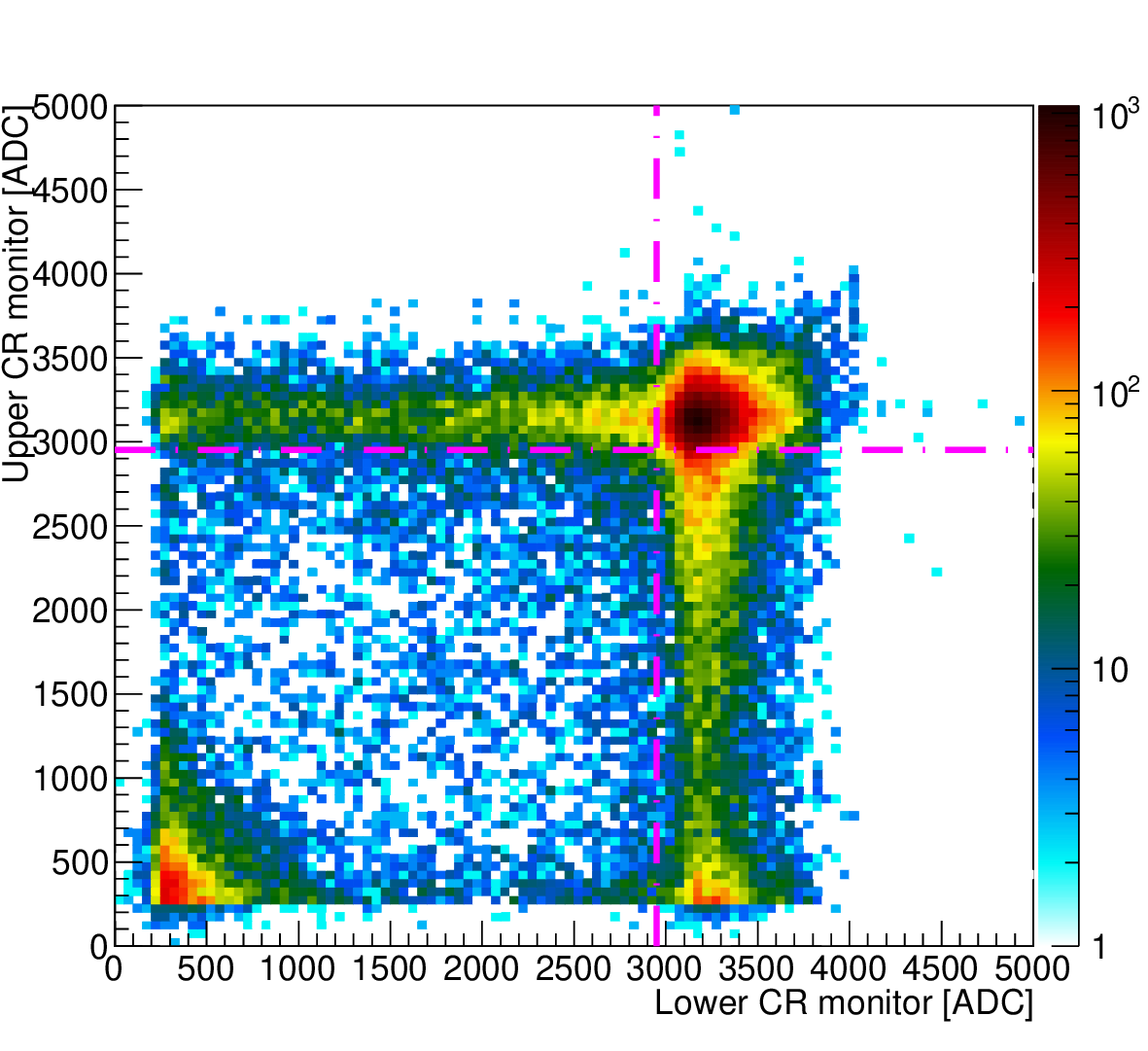}
\caption{Two-dimensional scatter diagrams of energy spectrum from two CR monitors.}
\label{fig:EffeCut}
\end{center}
\end{figure}

\begin{figure}
   \subfigure[]{
   \begin{minipage}[t]{0.95\linewidth}
   \centering
   \includegraphics[width=0.9\linewidth]{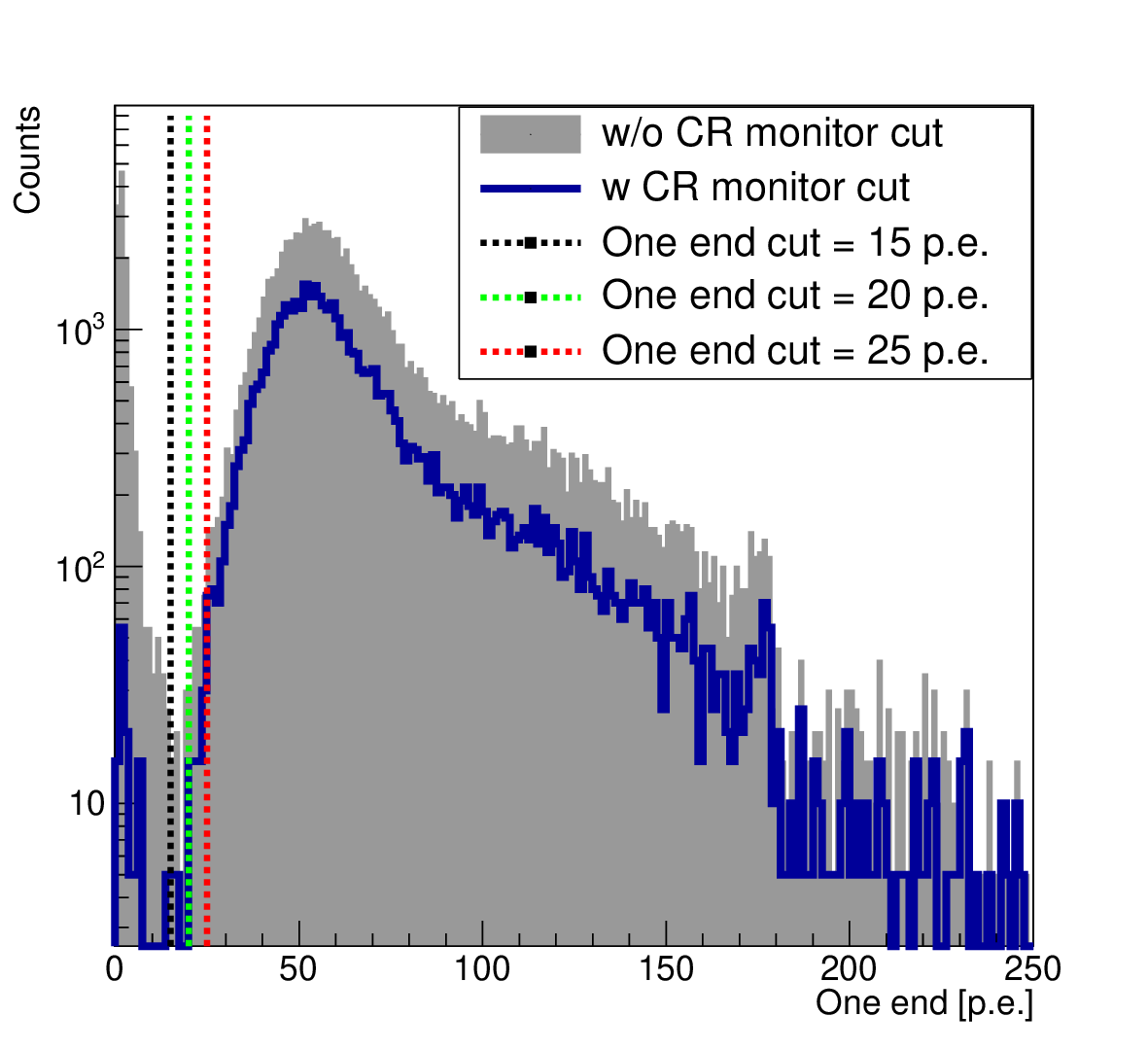}
    \label{fig:oneEff}
    \end{minipage}
    }
    \subfigure[]{
    \begin{minipage}[t]{0.95\linewidth}
    \centering
    \includegraphics[width=0.9\linewidth]{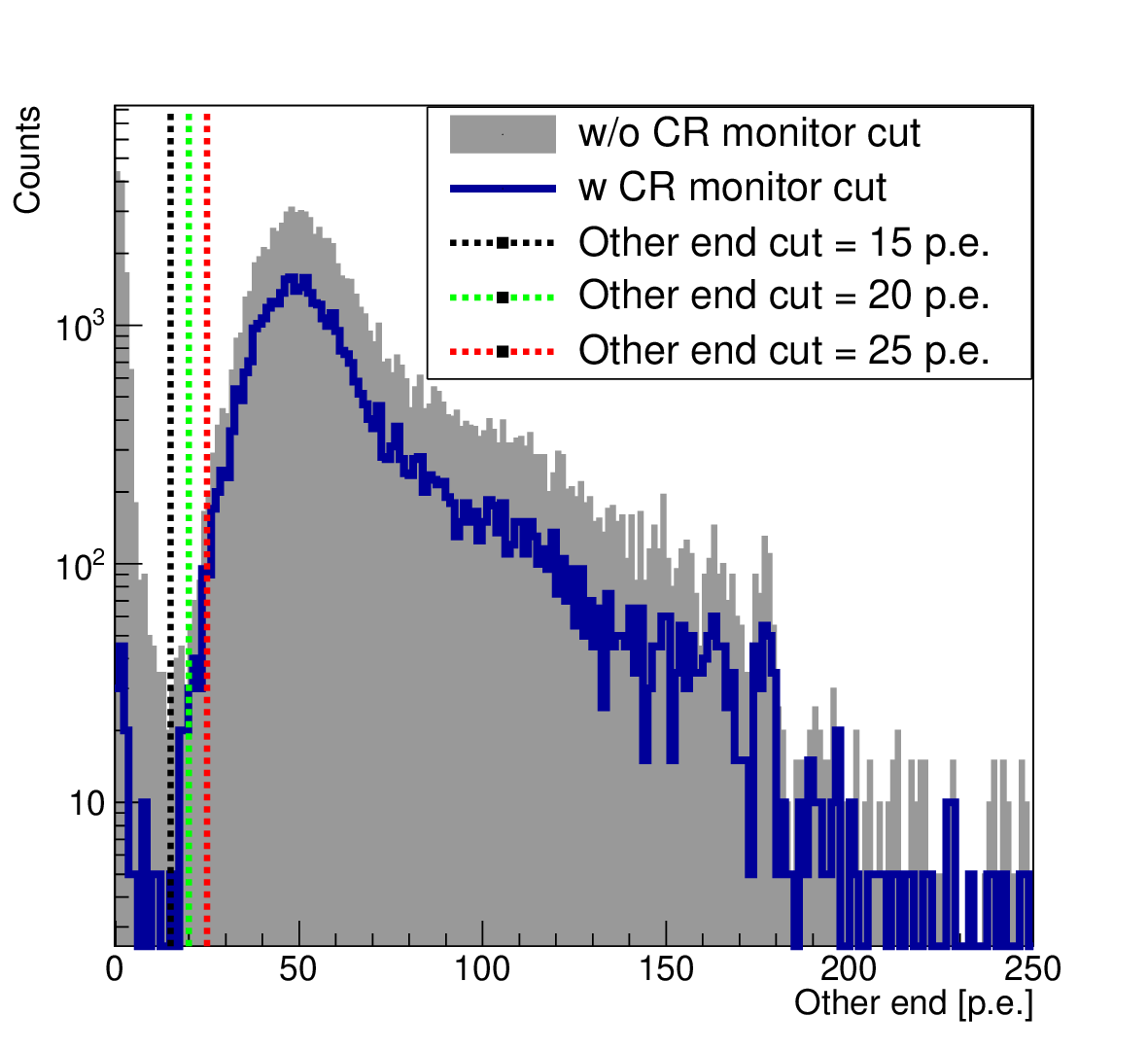}
    \label{fig:otherEff}
    \end{minipage}
    }
    \subfigure[]{
    \begin{minipage}[t]{0.95\linewidth}
    \centering
    \includegraphics[width=0.9\linewidth]{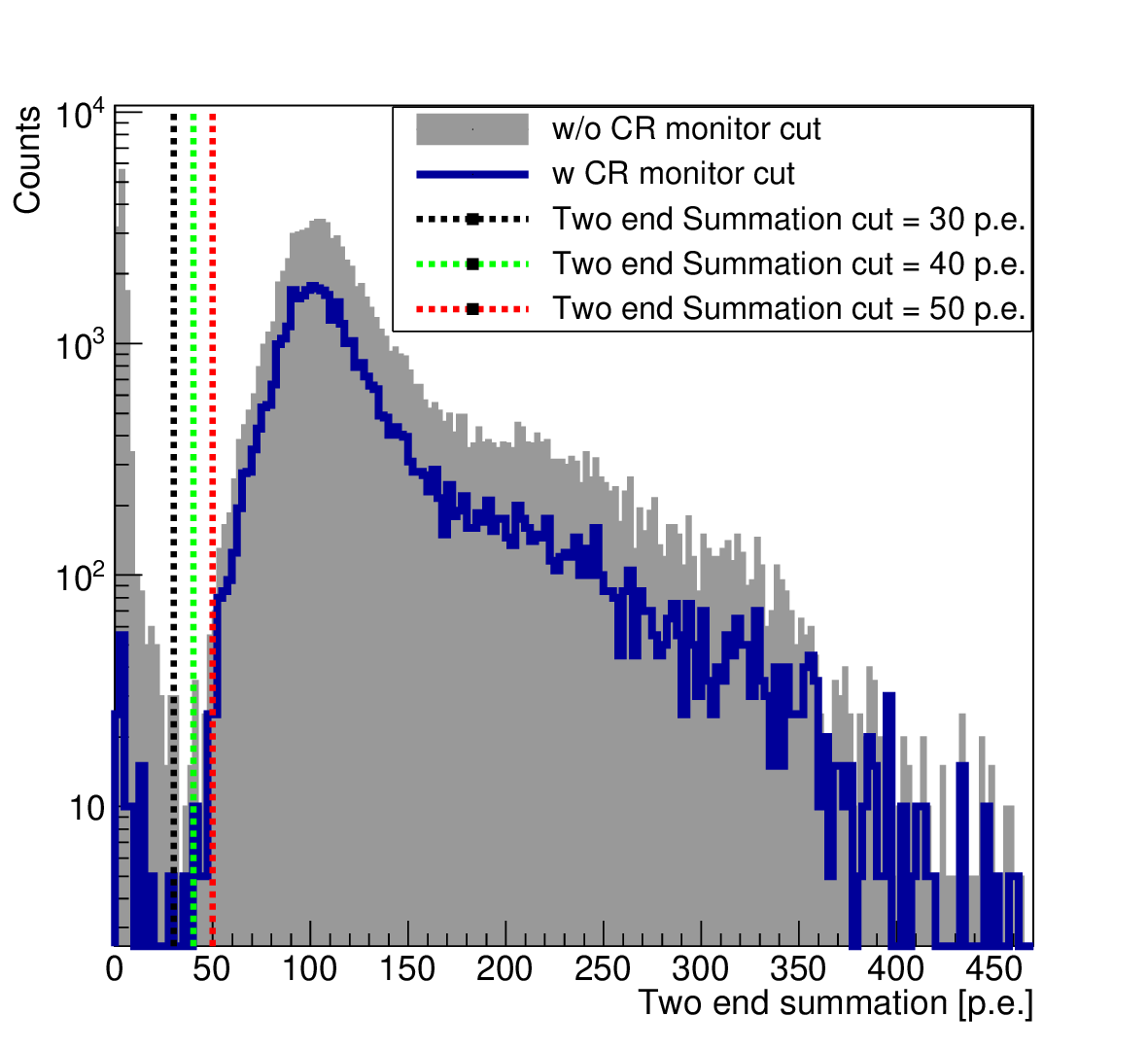}
    \label{fig:AllSumEff}
    \end{minipage}
    }
    \caption{(a) The charge spectrum of the summation channel at one end. (b) The charge spectrum of summation channel at the other end. (c) The sum of the two ends. }
    \label{fig:Eff}
\end{figure}

\begin{figure}[!htbp]
\begin{center}
\includegraphics[scale=0.42]{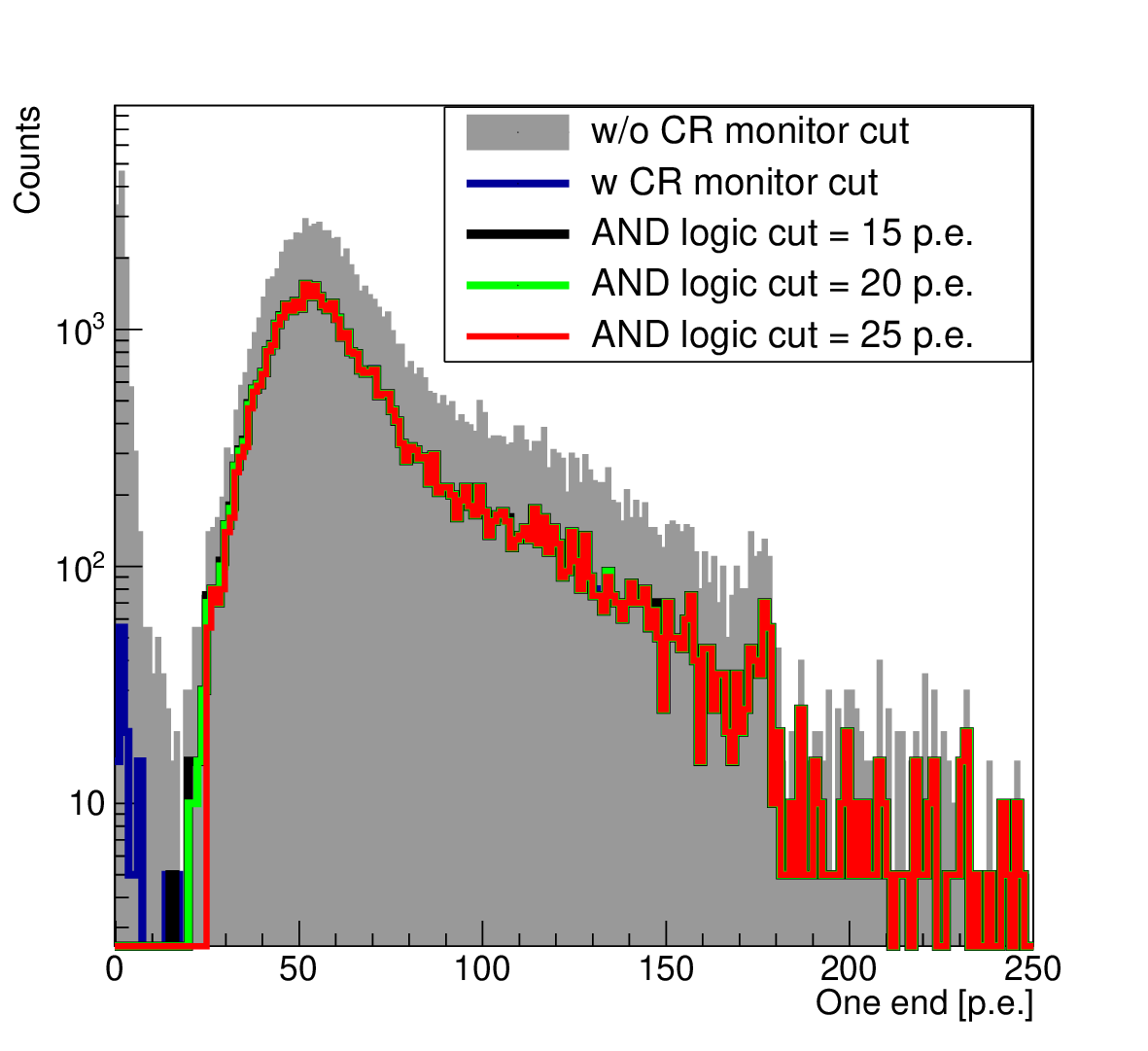}
\caption{The charge spectrum of the summation channel at one end when using an "AND" logic of the two ends of the PS module.}
\label{fig:AndLogic}
\end{center}
\end{figure}

Muons, passing through the two CR monitors in coincidence, could have passed through the PS module according to the setup. Further, the sample purity of muon candidates of the CR monitors can be tuned by the threshold of the CR monitor. 
Fig.\,\ref{fig:EffeCut} is a 2-D plot of the two CR monitors’ signals amplitude when it is placed at the center of the PS module. From the graph, it can be seen that there are several clusters, such as the cluster around 300\,ADC or 3100\,ADC. It is known that the cluster around 300\,ADC is mainly from the threshold effect and the environmental background, while the cluster around 3100\,ADC is mostly from our aimed CR muon. The shaded area represents the energy spectrum of all in Fig.\,\ref{fig:Eff}. 
The events in the upper right area of Fig.\,\ref{fig:EffeCut}, above 2940 (x) and 2940 (y) ADC of CR monitors, respectively, are considered as pure muon (we use $N_{reg}$ to represent the number of these events). 

With the further muon selection of the two CR monitors (above 2940 (x) and 2940 (y) ADC of CR monitors), we can obtain the signal strength of SiPM output mostly from muon from both ends of the PS strip. The blue plot in Fig.\,\ref{fig:oneEff} shows the charge spectrum of the summation channel of the four SiPMs at one end of the 1500mm-PS module, and the blue plot in Fig.\,\ref{fig:otherEff} is the summation channel of the other end. Fig.\,\ref{fig:AllSumEff} shows the total sum in p.e.\,of both ends from all eight SiPMs.

We checked three different thresholds (15\,p.e., 20\,p.e., 25\,p.e.) to one/another end (corresponding to the three different colors in Fig.\,\ref{fig:oneEff} or Fig.\,\ref{fig:otherEff}) to select detected muon events by one/another end, where we use $N[one]_{det}$ and $N[other]_{det} $ to represent the number of detected muons by one and another, respectively. In addition, we also checked the coincidence efficiency of different thresholds on both ends of the PS module, and the efficiency of different thresholds on the sum of both ends of the PS module considering the depression of background events. An "AND" logic of the two ends of the PS module for coincidence checking is used to select events (each end of the PS module is required above the threshold simultaneously in a time window 100\,ns), where we use $N[oneANDother]_{det}$ to represent the number of these events. 
Three thresholds (30\,p.e., 40\,p.e., 50\,p.e.) on the sum of both ends are checked (corresponding to the three different colors in the Fig.\,\ref{fig:AllSumEff}), where we use $N[module]_{det} $ to represent the number of these events.
Based on the strategy, we calculated the efficiency on different thresholds and configurations using the following four equations.

\begin{equation}
\epsilon_{one} = \frac{N[one]_{det}}{N_{reg}};
\label{equ:1}
\end{equation}

\begin{equation}
\epsilon_{other} = \frac{N[other]_{det}}{N_{reg}}
\label{equ:2}
\end{equation}

\begin{equation}
\epsilon_{module} = \frac{N[module]_{det}}{N_{reg}}
\label{equ:3}
\end{equation}

\begin{equation}
\epsilon_{oneANDother} = \frac{N[oneANDother]_{det}}{N_{reg}}
\label{equ:4}
\end{equation}

\begin{table}[!htb]
\centering
\caption{The calculated efficiency for each single end, the sum of the entire module, and the "AND" logic of the two ends with different thresholds.}
\label{table:DetEff}
\begin{tabular}{|c|c|c|c|} 
 \hline
 Threshold[p.e.]  & 15\,p.e. & 20\,p.e.& 25\,p.e.\\
 \hline
 $\epsilon_{one}$(\%)  & $99.72\pm 0.03$ & $99.69\pm 0.03$ & $99.53\pm 0.03$ \\
 \hline
 $\epsilon_{other}$(\%) & $99.70\pm 0.03$  & $99.62\pm 0.03$  & $99.46\pm 0.03$ \\
 \hline
 $\epsilon_{oneANDother}$(\%) & $99.63\pm 0.03$  & $99.48\pm 0.03$  & $99.01\pm 0.04$ \\
 \hline
 Threshold[p.e.] & 30\,p.e. & 40\,p.e.& 50\,p.e.\\
 \hline
 $\epsilon_{module}$(\%) & $99.70\pm 0.03$& $99.69\pm 0.03$ & $99.62\pm 0.03$ \\
 \hline
\end{tabular}
\end{table}

Tab.\,\ref{table:DetEff} shows the results of calculated detection efficiency with different thresholds on SiPM signal and configurations. From the overall perspective of the Tab.\,\ref{table:DetEff}, it can be observed that as the threshold of the SiPM signal increases, the detection efficiency gradually decreases as expected. From Tab.\,\ref{table:DetEff}, it can also be observed that $\epsilon_{oneANDother}$ decreases the fastest as the threshold increases. Fig.\,\ref{fig:AndLogic} displayed the energy spectrum under "AND" logic mode. It can be observed that in this mode, there is a significant decrease in the coincidence background. 

\begin{table}[!htb]
\centering
\caption{The rate vs. threshold of PS module with sum of both ends, and the coincidence of both ends using "AND" logic.}
\label{table:Detrate}
\begin{tabular}{|c|c|c|c|} 
 \hline
 Threshold[p.e.] & 30\,p.e. & 40\,p.e.& 50\,p.e.\\
 \hline
 $Rate_{module}$(Hz) & $93\pm 9.7$& $83\pm 9.1$ & $70\pm 8.3$ \\
 \hline
 Threshold[p.e.]  & 15\,p.e. & 20\,p.e.& 25\,p.e.\\
 \hline
 $Rate_{oneANDother}$(Hz) & $73\pm 8.5$  & $63\pm 7.9$  & $48\pm 6.9$ \\
 \hline
\end{tabular}
\end{table}

Tab.\,\ref{table:Detrate} shows the rate vs.\,threshold of the sum of both ends of the PS module, and the coincidence of the two ends when the PS module located at the laboratory on ground without CR monitor. From Tab.\,\ref{table:Detrate}, it can also be observed that under this "AND" logic mode, the events of backgrounds that coincidentally match sharply decreases. At the same time, it can be found that the rate(63\,Hz) corresponding to the threshold of 20\,p.e. is close to the muon rate(64\,Hz) at sea level.

\begin{figure}[!htbp]
\begin{center}
\includegraphics[scale=0.42]{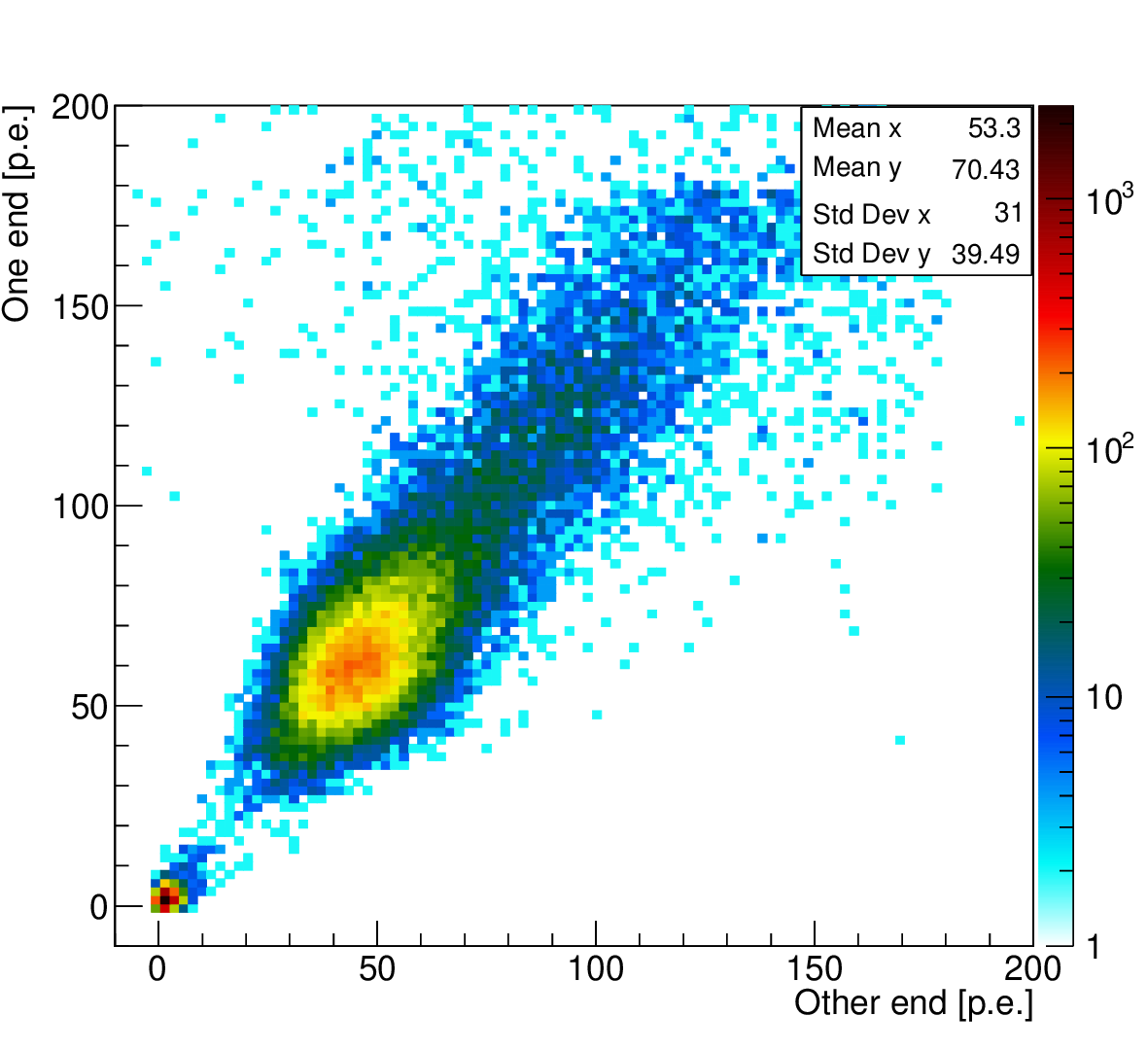}
\caption{2-D plot of the two ends when the CR monitors are at 60\,cm away from the center of the PS module.}
\label{fig:Assymtry}
\end{center}
\end{figure}

\begin{figure}
   \subfigure[]{
   \begin{minipage}[t]{0.98\linewidth}
   \centering
   \includegraphics[width=0.80\linewidth]{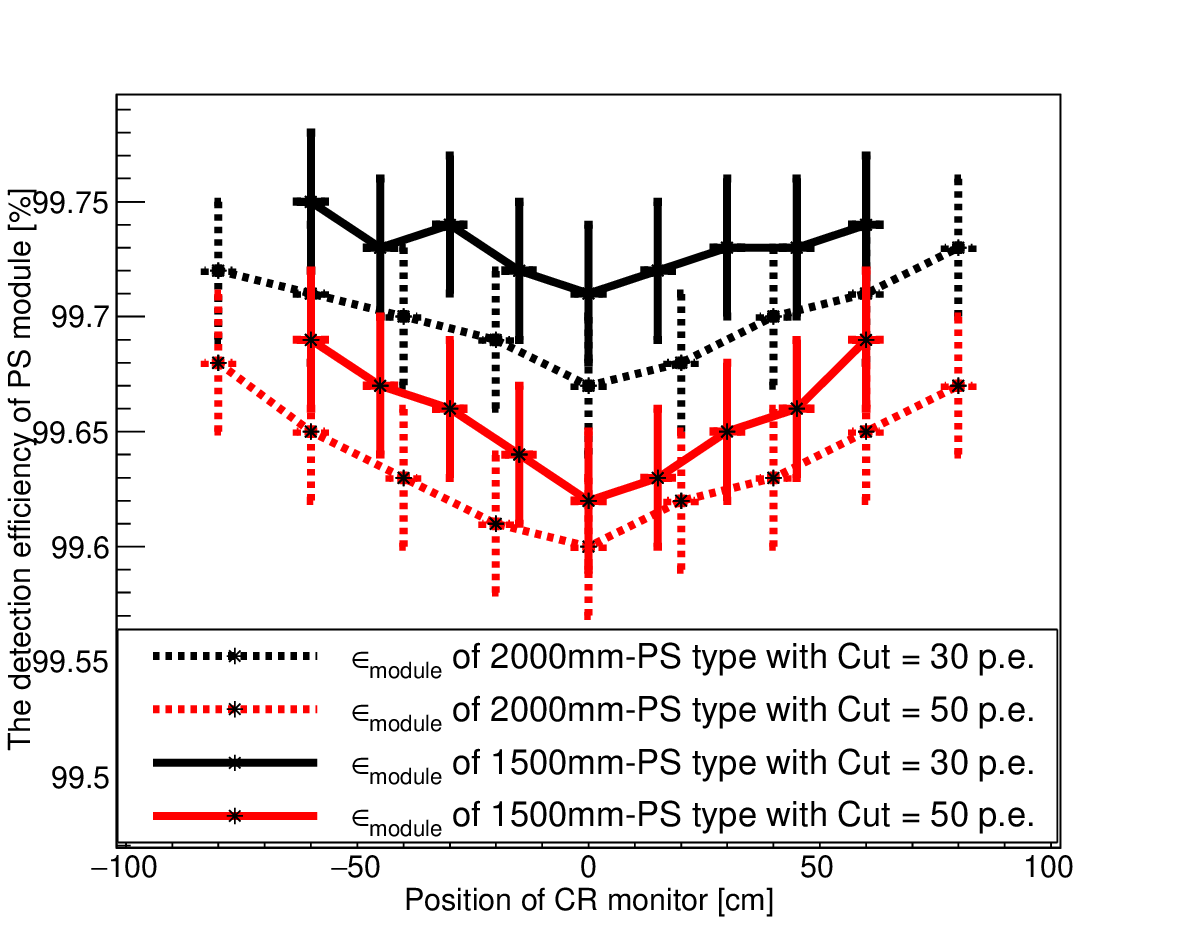}
    \label{fig:Effdisf}
    \end{minipage}
    }
    \subfigure[]{
    \begin{minipage}[t]{0.98\linewidth}
    \centering
    \includegraphics[width=0.80\linewidth]{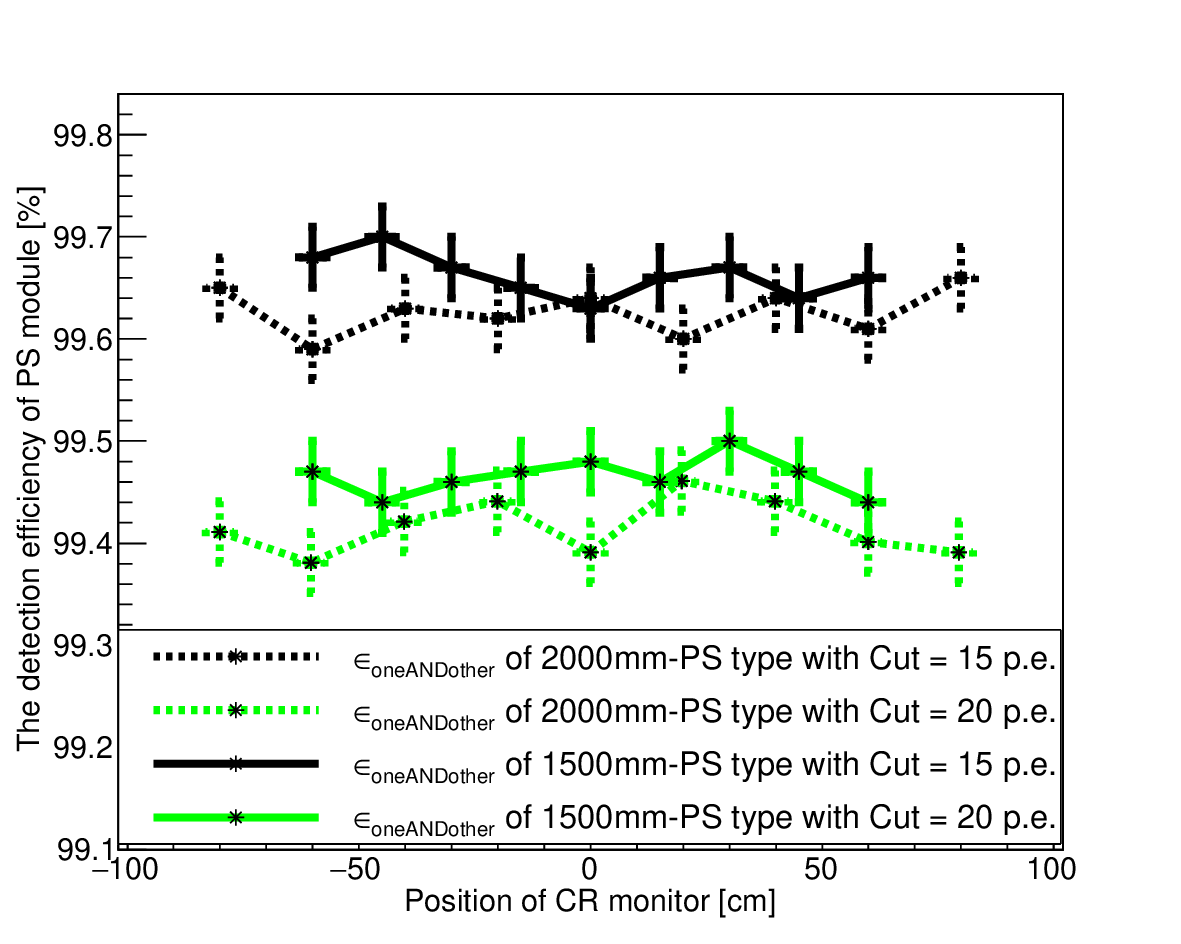}
    \label{fig:Efforand}
    \end{minipage}
    }
    \caption{(a) The efficiency of different sum thresholds of the 1500\,mm and 2000\,mm PS modules. (b) The efficiency with different "AND" logic thresholds for the 1500\,mm and 2000\,mm PS modules.}
\end{figure}

The CR monitors were also placed at other positions of the PS modules to check the efficiency. 
When the CR monitors are off to the center of the PS module, the signal strength obtained at the two ends of the PS module will be different. 
Fig.\,\ref{fig:Assymtry} shows the 2-D plot of the two ends of the PS module when the muon hits around 60\,cm away from the center of the PS module. It can be clearly seen that the noise or environmental background can be identified from the signal of the muon. But the signal at one end is significantly larger than the other end, which could generate a difference in the efficiency of the two ends with the same threshold on SiPM signals. The muon signal exhibits elliptical characteristics on the two-dimensional graph, which is due to the asymmetric distance of the CR monitor from the center of PS strip. Combining Fig.\,\ref{fig:RestS5} and Fig.\,\ref{fig:RestS6}, as well as the two-dimensional graph in Fig.\,\ref{fig:Assymtry}, they all indicated that the smaller the MPV of the Landau distribution, the smaller the width of its Landau distribution. When the threshold is too high, even approaching the MPV of one end, the $\epsilon_{oneANDother}$ be determined by the far end (i.e. the end with the smaller signal). Through the measurement results of CR monitor at the outermost edge of PS strip, we found that when the threshold is 15 or 20\, p.e., the $\epsilon_{oneANDother}$ is still not affected, almost equivalent to the efficiency at the center. When the threshold is greater than 22\, p.e., the $\epsilon_{oneANDother}$ at the edge was approximately 0.3\% lower than at the center. By combining with Tab.\,\ref{table:Detrate}, it is demonstrated that a very good tolerance on the threshold to reach the muon tagging efficiency is possessed by us.
Considering the possible efficiency and background under different thresholds, we ultimately decided to use $\epsilon_{module}$ and $\epsilon_{oneANDother}$ as the characterization of overall detection efficiency of PS module.

Fig.\,\ref{fig:Effdisf} shows the relationship between detection efficiency of the sum of the two ends and muon position (CR monitors' position). The position measurements were made by moving the monitors along the length by a 20\,cm step for the 2000mm-PS module (indicated in the figure by dotted lines), or by a 15\,cm step for the 1500mm-PS module (indicated in the figure by solid lines). The error bar of the X represents the 6\,cm size width of the CR monitor, the Y-coordinate is the detection efficiency of the PS module when the threshold of the sum of the two ends is set to 30\,p.e.\,or 50\,p.e., the error bar of the Y-axis is the error in efficiency. As the threshold increases, the detection efficiency corresponding to all positions of the PS module decreases. For a 1500mm-PS module, when the threshold is set to 30\,p.e., the detection efficiency of each point is higher than 99.7\%. For a 2000mm-PS module, when the threshold is set to 50\,p.e., the detection efficiency of each point is higher than 99.6\%. Overall, the 1500-mm PS detection efficiency is higher than that of 2000-mm PS. 
Fig.\,\ref{fig:Efforand} shows the relationship between detection efficiency of "AND" logic of the two ends and muon position. In this mode, the efficiency of 1500mm-PS is slightly better than that of 2000mm-PS. For all PS module, when the threshold is set to 20 p.e., the detection efficiency of each point is better
than 99.3\%. The efficiency shows almost no difference along the entire module at 0.1\% level.

\section{Conclusion}
\label{sec:Sum}
In this study, we provided a detailed introduction to the unique design and performance of the module, which holds meaningful value for the process design of plastic scintillator detectors with WLS-fibers. Additionally, we proposed a batch quality inspection process and established a set of standards for evaluating the performance of plastic scintillator modules, offering valuable experience and reference for quality inspection in other related experiments.
The specific summary is as follows:
The more the muon hits the position towards the two ends of the PS strip, the higher the total light yield but with a stronger asymmetry. When the muon hits the center of the PS strip (minimum effective light yield region along the strip, air coupling between SiPM and PS), for a 2000mm-PS module, the most probable signal strength output from one end is at least greater than 40.8\,p.e. For a 1500mm-PS module, it is at least greater than 51.5\,p.e. All the modules meet the requirement proposed by JUNO-TAO that the light yield to a muon should be greater than 40\,p.e. With optical grease to couple SiPM and WLS-fiber, it can increase the effective light yield by 12.5\%. 

The muon detection efficiency is checked with different thresholds on the output of single end, sum or coincidence of two ends of the module. Three types of detection efficiency have been defined by us, comprehensively evaluating the performance of the plastic scintillator. Overall, the 1500-mm PS detection efficiency is better than that of 2000-mm PS. In "AND" mode, when the threshold is set at 20 photoelectrons, the detection efficiency exceeds 99.3\%, and at 15 photoelectrons, it surpasses 99.6\%, with clear differentiation between background and muon signals. These features and advantages result in our plastic scintillator performing exceptionally well in experiments, providing a reliable detection solution for other experiments.
\section{Acknowledgements}
\label{sec:ACK}
This work is supported by the General Program of National Natural Science Foundation of China (Project No.12075087), the National Key Research and Development Program of China (Project No.2022YFA1602002), the Strategic Priority Research Program of the Chinese Academy of Sciences (Project No. XDA10011102). Guang Luo also greatly appreciates the support from Guangdong Provincial Key Laboratory of Advanced Particle Detection Technology. 

\end{document}